\documentclass{article}
\usepackage{jheppub}
\usepackage[utf8]{inputenc}
\usepackage{amsmath}
\usepackage{amssymb}
\usepackage{dsfont}
\usepackage{tensor}
\usepackage{mathtools}
\usepackage[makeroom]{cancel}
\usepackage{dsfont}
\usepackage{hyperref}
\hypersetup{
    colorlinks=true,
    linkcolor= black,
    filecolor=magenta,      
    urlcolor=cyan,
    pdftitle={Overleaf Example},
    pdfpagemode=FullScreen,
    }
\usepackage{textgreek}
\usepackage{simpler-wick}
\usepackage{tikz-cd}

\newcommand{\Tr}{\text{Tr}\,}
\newcommand*\dd{\mathop{}\!\mathrm{d}}

\newcommand{\dbar}{\bar{\partial}}
\newcommand{\psibar}{\bar{\psi}}
\newcommand{\bra}{\left\langle}
\newcommand{\ket}{\right\rangle}
\newcommand\bC{\mathbb{C}}
\newcommand\bZ{\mathbb{Z}}
\newcommand\bR{\mathbb{R}}
\newcommand\bP{\mathbb{P}\,}
\newcommand{\tsr}[2]{\tensor{#1}{#2}}

\newcommand\1{{\mathds{1}}}
\newcommand{\pf}{\text{Pf}}
\title{D-Brane Systems in Twisted Holography and \texorpdfstring{$SO$/$Sp$}{SO/Sp}  Chiral Algebras}
\author[1,2]{Adri\'an L\'opez-Raven}
\emailAdd{alopezraven@perimeterinstitute.ca}
\affiliation[1]{Perimeter Institute for Theoretical Physics, Waterloo, ON N2L 2Y5, Canada}
\affiliation[2]{Department of Physics \& Astronomy, University of Waterloo, Waterloo, ON N2L 3G1,
Canada}
\date{\today}
\abstract{
We study correlation functions of baryon and determinant operators for the chiral algebras obtained from the twist of $\mathcal{N}=4$ SYM with $U(N)$ gauge group. In the context of Twisted Holography, we conjecture that a dual description should involve a D1-D5 brane system, and we construct from the correlators a candidate dual brane in the form of a derived coherent sheaf in $SL_2(\bC)$. Extending this analysis, we compute similar baryon/determinant correlators of chiral algebras in symmetric and antisymmetric representations of $SO$ and $Sp$ gauge groups and construct the candidate dual branes. These branes exhibit $\mathbb{Z}_2$ identifications consistent with conjectures relating $SO/Sp$ chiral algebras to Kodaira-Spencer theory on $SL_2(\mathbb{C})/\mathbb{Z}_2$ and the Type I topological string on $SL_2(\mathbb{C})$.
}

\begin{document}

\maketitle
\section{Introduction}

In 2018, Costello and Gaiotto introduced a novel holographic correspondence known as Twisted Holography. This duality, already foreshadowed in \cite{Gopakumar:1998ki} as a B-model analog to Gopakumar-Vafa duality, and partially explored in \cite{Bonetti:2016nma}, was formulated precisely in \cite{costello2021twistedholography}\footnote{A holographic setup similar in spirit to \cite{costello2021twistedholography} can also be found in \cite{Ishtiaque:2018str_topological_holography}.}. A concise summary of the topic can be found in \cite{Budzik:2023nnx_non_conformal_vacua}, which highlights how Twisted Holography emerges as the supersymmetric twist of the standard Maldacena holography.  

Twisted Holography can also be understood as the application of a Maldacena-type construction to the B-model. Specifically, one considers a stack of \( N \) D-branes placed along a \( \mathbb{C} \) subspace of a \( \mathbb{C}^3 \) bulk. In the large-\( N \) limit, the worldvolume theory on the D-brane stack is found to be holographically dual to the bulk gravity theory in a backreacted background. For the B-model, the stack supports a two-dimensional chiral algebra as its worldvolume theory. On the other hand, the bulk theory corresponds to Kodaira-Spencer theory \cite{Bershadsky:1993cx_Kodaira_Spencer_Vafa}, also known as BCOV theory \cite{costello2012quantumbcovtheorycalabiyau}, defined in a backreacted geometry that transitions from $\mathbb{C}^3$ to $SL_2(\mathbb{C})$.\\

% Twisted Holography, as they came to call it, is to the B-model what the standard Maldacena
% holography is to the type IIB string. That is, in both cases you place a stack of N D-branes and in
% the large N limit, one finds that the worldvolume theory on the stack is holographically dual to the
% gravity theory in a backreacted background. In the case of the B-model, one places a stack D1-branes
% supported on a C inside a C3 bulk. The stack supports a 2d chiral algebra as a worldvolume theory.
% The bulk theory instead is Kodaira Spencer, also called BCOV theory, [5] in a background which after
% backreaction goes form the C3 to an SL2C geometry.

% We'll review the chiral algebras in detail in Section \ref{sec:gauged_chiral_algebras}, here we briefly mention that they're made of (some BRST cohomology of) two matrix-valued fields, $X(z), Y(z)$ living on the complex plane, valued in the adjoint, symmetric or antisymmetric representation of some gauge group.
% \begin{align}
% X(z), Y(z)
% \end{align}

In this work we propose a novel entry in the Twisted Holography correspondence, matching baryonic operators in the boundary chiral algebra and D1-D5 brane systems in the bulk. \\

A family of D1 branes was found in \cite{giant_gravitons_kasia}, arising in the bulk from the insertion of determinant operators in the boundary. More specifically, starting from determinant insertions in the boundary chiral algebra, they identified the surface in the bulk wrapped by the dual D1 brane and the associated Chan-Paton bundle.

As for the space-filling D5's, they were originally introduced in \cite{costello2021twistedholography} and conjectured to be in correspondence with the introduction of additional boundary ``quarks", $J(z), I(z)$, living in the fundamental and antifundamental representation of the gauge group.

This dictionary
\begin{align}
\det(\mathcal{O})   &\xleftrightarrow{\qquad} \text{D1-Branes}\\
I(z),J(z) &\xleftrightarrow{\qquad} \text{D5-Branes}
\end{align}
motivated the driving question of this work: \emph{In the presence of both determinant operators and quarks, can one find novel instantonic deformations like those found in the physical theories in Dp-D(p-4) brane systems \cite{Witten:1994tz_instantons_D_brane_systems}\cite{Douglas:1996uz_gauge_fields_and_d_branes}\cite{Douglas:1995bn_branes_with_branes}? }

The answer is in the affirmative: By recasting the  determinants as integrals over 0d fermionic fields
\begin{align}
\det(\mathcal{O}) = \int [d\psi d\psibar] e^{\psibar \mathcal{O}\psi}
\end{align}

one may add an additional deformation mixing both the $I,J$ fields with $\psi$ and $\bar\psi$
\begin{align}
\int [d\psi d\psibar] e^{\psibar \mathcal{O}\psi +  I\psi + \bar\psi J}
\end{align}
For these operators, we are able to construct a candidate dual brane system, in the form of a derived coherent sheaf. That is, we construct a chain complex that encodes both the data of the deformed space wrapped by the brane and the associated Chan-Paton bundle\footnote{We make brief remarks of why these complexes encode the data of a brane in Section \ref{sec:a_brane_as_a_complex}. A detailed introduction to branes as derived coherent sheaves can be found in \cite{Sharpe:2003dr_D_branes_and_sheaves} and \cite{Pietromonaco:2017xup_derived_category_of_coherent_sheaves}.}. Moreover, our construction subsumes the giant graviton construction from \cite{giant_gravitons_kasia} as a subcase, and provides for it a more symmetric presentation.
% that doesn't single out the $a$ (as in the 1,1 component in ${a\,\, b \choose c\,\, d} \in SL_2(\bC)$) coordinate
% To this new operator, we are able to assign a a derived coherent sheaf that represents the deformed dual brane system. That is, both the surface of the brane, and the associated Chan-Paton bundle are provided in terms of the cohomology of some chain complex\footnote{We make some remarks of why these complexes encode the data of a brane in Section \ref{sec:a_brane_as_a_complex}. A detailed introduction to branes as derived coherent sheaves can be found in \cite{Sharpe:2003dr_D_branes_and_sheaves} and \cite{Pietromonaco:2017xup_derived_category_of_coherent_sheaves}}. Moreover, our construction subsumes the giant graviton construction from \cite{giant_gravitons_kasia} as a subcase, and provides for it a more symmetric presentation.

This work arose from an effort to identify instantonic deformations in the chiral algebra where matter fields are in the antisymmetric representation of an \( Sp(2N) \) gauge group. The question naturally emerged as the presence of quarks is required to avoid gauge anomalies. 

% These BRST anomalies find a nice correspondence with what is conjectured to be the holographic dual to this chiral algebra \cite{Jake_Abajian:2024rjq}: the Type I Topological String defined in \cite{costello2020anomalycancellationtopologicalstring}. In the bulk, anomaly cancellation via a Green-Schwarz mechanism requires the introduction of eight space-filling branes. This number precisely matches the eight \( I(z) \) fields needed to cancel the BRST anomaly in the boundary chiral algebra.

The second half of this work studies the instanton deformations for this chiral algebra and other cases involving symmetric and antisymmetric representations of both gauge groups $Sp$ and $SO$. In all cases, we construct the candidate dual branes as a derived coherent sheaf. In contrast with the $U(N)$ chiral algebra, these branes exhibit identifications between the Chan-Paton bundle and its dual bundle, in addition to $\bZ_2$ identifications relating different points of the brane. These findings provide evidence for conjectures in the literature \cite{Jake_Abajian:2024rjq} relating these $SO/Sp$ chiral algebras to Kodaira-Spencer theory on $SL_2(\mathbb{C})/\mathbb{Z}_2$ and the Type I Topological String \cite{costello2020anomalycancellationtopologicalstring} on $SL_2(\mathbb{C})$.

\subsection{Structure of the Paper}

% In Section \ref{sec:On_the_Bulk_Geometry_and_its_Symmetries}, we review how to construct the asymptotic boundary of $SL_2(\bC)$ and mention some of its symmetries. In Section \ref{sec:gauged_chiral_algebras}, we review the definition of the $U(N)$ chiral algebra involved in twisted holography, and address some aspects of its BRST cohomology. In Section \ref{sec:A_Complex_for_an_Instanton}, we define the brayon/determinant operators for which we conjecture a brane dual and provide the main constructions of this work: the chain complexes associated to the different dual branes. In Section \ref{sec:SO_Sp} we define the chiral algebras with $SO(N)$ and $Sp(2N)$ gauge group, and construct the corresponding complexes for the dual branes. We further present how these branes possess $\bZ_2$ identifications that provide evidence for conjectures regarding the holographic duals to these $SO$/$Sp$ chiral algebras.

In Section \ref{sec:On_the_Bulk_Geometry_and_its_Symmetries}, we review the construction of the asymptotic boundary of $SL_2(\bC)$ and discuss some of its symmetries. Section \ref{sec:gauged_chiral_algebras} introduces the $U(N)$ chiral algebra relevant for twisted holography and examines certain aspects of its BRST cohomology. In Section \ref{sec:A_Complex_for_an_Instanton}, we define the brayon and determinant operators for which we propose a conjectural dual brane system, and present the main construction of this work: the dual brane in the form of a derived coherent sheaf. Section \ref{sec:SO_Sp} introduces the chiral algebras with $SO(N)$ and $Sp(2N)$ gauge groups and constructs the corresponding complexes for their dual branes. Additionally, we demonstrate how these branes exhibit $\bZ_2$ identifications, providing evidence for conjectures about the holographic duals of the $SO$ and $Sp$ chiral algebras.

\section{On the Bulk Geometry and its Symmetries}\label{sec:On_the_Bulk_Geometry_and_its_Symmetries}
All the chiral algebras we study in this work have as their holographic dual the Kodaira-Spencer theory on an \( SL_2(\mathbb{C}) \) background (or a related orientifold theory). However, in this work we do not make use of specific details of Kodaira-Spencer theory. Instead, we focus on reviewing key properties of the underlying \( SL_2(\mathbb{C}) \) geometry, which facilitate elementary holographic matches with the boundary chiral algebra.  

% All the chiral algebras we'll study will have as holographic \dg{typo} dual the Kodaira-Spencer Theory on an $SL_2(\bC)$ (or some orientifold theorof) background. We'll have nothing to say about Kodaira-Spencer in this work. Instead, we review some properties of the underlying $SL_2(\bC)$ geometry that provide some elementary holograhic matches to the boundary chiral algebra. \al{improve} \dg{Improve presentation}

First, we review how to construct the conformal boundary of $SL_2(\bC)$. $SL_2(\bC)$ is defined as the space of matrices that satisfy the relation:
\begin{align}
ad - bc = 1
\end{align}

% All the chiral algebra we'll study have two $SL(2)$ symmetries we'll make frequent use of \al{maybe change sentence}: Global conformal transformations and an $SL(2)$ R-symmetry\footnote{This symmetry is a remnant of the R-symmetry of the original untwisted theory}  mixing $X$ and $Y$. These two find a natural holographic correspondence, respectively, on the left and right actions of $SL_2(\bC)$ on itself, which constitute symmetries of the bulk theory. Following \cite{costello2021twistedholography}, we refer to these as $SL_L(2)$ and $SL_R(2)$.

To construct its conformal boundary, we apply analogous heuristics to those used for defining the boundary of \( AdS \) \cite{witten1998antisitterspaceholography}. To make \( a \), \( b \), \( c \), and \( d \) approach infinity, we scale them simultaneously by a constant factor \( k \):
\begin{align}
k^2(ad - bc) = 1
\end{align}

and then take the limit as \( k \to \infty \). In this limit, the points at infinity must satisfy the condition
\begin{align}\label{eq:conformal_boundary_equation}
ad - bc = 0
\end{align}

where, having allowed ourselves an arbitrary scaling in the limit to find \eqref{eq:conformal_boundary_equation}, we take it that $a,b,c,d$ are defined up to a simultaneous rescaling\footnote{This approach can be made more rigorous, as discussed in section 7 of \cite{costello2021twistedholography}, by noting that a natural compactification of \( SL_2(\mathbb{C}) \) is the projective variety \( \{(A:B:C:D:V) \mid AD - BC = V^2 \} \). The original \( SL_2(\mathbb{C}) \) corresponds to the patch \( V \neq 0 \), while the points with \( V = 0 \), i.e., \( \{(A:B:C:D) \mid AD - BC = 0 \} \), define the conformal boundary.}. Thus we conclude, that the conformal boundary of $SL_2(\bC)$ consists of matrices of rank 1 defined up to scaling. Additionally, these matrices can be factorized
\begin{align}
\begin{pmatrix}
a & b\\
c & d
\end{pmatrix}=
\begin{pmatrix}
q_1\\
q_2
\end{pmatrix}(p_1 \quad p_2)
\end{align}

where each factor is separately defined up to scaling. This leads to the punchline
\begin{equation}
\partial SL_2(\bC) \cong (\bC^2 \times \bC^2) / \text{Scale} \cong \bC P^1\times\bC P^1
\end{equation}

The chiral algebra we introduce in the next section consist of (forgetting ghosts) two fields $X(z)$ and $Y(z)$. The first $\bC P^1$ is their base space, while the second is their target. 

To parametrize the $\bC P^1$'s, we use the coordinates
\begin{align}\label{eq:boundary_CP1_coordinates}
z \coloneqq \frac{q_2}{q_1} = \frac{c}{a} = \frac{d}{b} \quad \text{and} \quad u \coloneqq \frac{p_2}{p_1} = \frac{b}{a} = \frac{d}{c}
\end{align}

Having described the boundary, we now highlight two symmetries of $SL_2(\bC)$ that find a natural match with symmetries of our chiral algebras.  $SL_2(\bC)$ may act on itself via group multiplication from the left and from the right. Following \cite{costello2021twistedholography}, we denote these two actions as $SL_L(2)$ and $SL_R(2)$. Their actions on bulk and boundary are:

For \( SL_L(2) \):
\begin{align}\label{eq:SL2L_definition}
\begin{pmatrix}
a & b\\
c & d
\end{pmatrix}\mapsto
\begin{pmatrix}
\alpha & \beta\\
\gamma & \delta
\end{pmatrix}
\begin{pmatrix}
a & b\\
c & d
\end{pmatrix};\quad z \mapsto \frac{\delta z + \gamma}{\beta z + \alpha};\quad
u \mapsto u
\end{align}    

For $SL_R(2)$: 
\begin{align}\label{eq:SL2R_definition}
\begin{pmatrix}
a & b\\
c & d
\end{pmatrix}\mapsto
\begin{pmatrix}
a & b\\
c & d
\end{pmatrix}
\begin{pmatrix}
\alpha & \beta\\
\gamma & \delta
\end{pmatrix};\quad
z \mapsto z;\quad
u \mapsto \frac{\delta u + \beta}{\gamma u + \alpha}
\end{align}
\section{Lightning Review of Gauged Chiral Algebras and their BRST Cohomology}\label{sec:gauged_chiral_algebras}
% The chiral algebras we'll consider consist of two matrix-valued fields,  $X(z),Y(z)$, with conformal weight $h=\frac{1}{2}$, with a gauged $U(N)$, $SO$ or $Sp$ symmetry. To fix ideas, in what follows we'll refer only to the $U(N)$ case, but similar considerations follow identically in all other cases.

Here we give a brief review on some properties of the chiral algebras we study in the bulk of this paper, and give some remarks about their BRST cohomologies.

Our chiral algebras will always consist of two matrix-valued fields,  $X(z),Y(z)$, with conformal weight $h=\frac{1}{2}$, with a gauged $U(N)$, $SO(N)$ or $Sp(N)$ symmetry. In the $U(N)$ case, $X$ and $Y$ will live in the adjoint representation, while in the $SO$ and $Sp$ chiral algebras of section \ref{sec:SO_Sp}, we'll study both the case when they live in the symmetric or antisymmetric reps for each gauge group.

In all these theories, up to normalization factors, the action is given by:
\begin{equation}
S = N\int \dd^2 z \,\Tr X\bar{D}Y% = \int \dd^2 z \,\Tr \frac{1}{2} \omega_{\alpha\beta}Z^\alpha\bar{D} Z^\beta
\end{equation}

% with $\omega = {\,\,\,\,0 \quad 1\choose -1 \quad 0}$  and 
where $\bar{D}$ is an anti-holomorphic covariant derivative $\bar{D} = \bar\partial + [\bar{A}, \,\cdot\,\,]$, and $\bar{A}$ the corresponding gauge field.

Since the gauge field has only one spacetime component, we may gauge fix it away (i.e use the gauge fixing condition $\bar{A}=0$), which after some BRST gymnastics, leads to the action: 
\begin{equation}\label{eq_ad:chiral algebra action}
S = N\int \dd^2 z \,\Tr X\dbar Y  + \Tr b\dbar c
\end{equation}

Note that at this point there is no interacting term in our theory, instead the only remnant of the interaction with $\bar A$ is that we must work in BRST cohomology. 

% Additionally, observe that the theory has an $SL(2,\bC)$. symmetry, which rotates $X$ and $Y$ \al{say something about $SL_{2R}$ and their holographic interpretation}:

% \begin{equation}
% \begin{pmatrix}
% X\\
% Y
% \end{pmatrix}
% \mapsto g 
% \begin{pmatrix}
% X\\
% Y
% \end{pmatrix}
% \text{with } g\in SL_2(\bC)
% \end{equation}

The OPEs of the $U(N)$ chiral algebra are given by:
\begin{align}\label{eq_ad:OPEs_U(N)}
\tsr{X}{^{a}_b}(z) \tsr{Y}{^{c}_d}(z') &\sim
\frac{1}{N}\dfrac{1}{z - z'}\; \delta^a_d \delta ^b_c  \\
\tsr{b}{^{a}_b}(z) \,\tsr{c}{^{c}_d}(z') &\sim
\frac{1}{N}\dfrac{1}{z - z'}\; \delta^a_d \delta ^b_c  
\end{align}

% where the $\hbar$ is introduced as a parameter that counts wick contractions. We'll sometimes use a 't Hooft parameter $ \lambda \coloneqq \hbar N $ as the wick contraction counting parameter so $\frac{1}{N}$ 

% The $\hbar$ parameter is introduced to be able to distinguish the classical limit ($\hbar \rightarrow 0 $ and $N$ fixed) from the 't Hooft limit ($\hbar \rightarrow 0$ and $N \rightarrow \infty$ with 't Hooft coupling $\lambda \coloneqq \hbar N$ fixed). We elaborate on this choice of scaling in Section \ref{sec:brief_remarks_on_factors_of_hbar}. \al{maybe elaborate}

Holographically, we interpret the $X,Y$ fields as living in the first boundary $\bC P^1$, and spanning the second. By the latter statement, we mean that given a linear combination
\begin{align}
p_1X + p_2 Y    
\end{align}

we interpret the $p_1,p_2$ coefficients as homogeneous coordinates of the second $\bC P^1$. In this way, one can define the operator
\begin{align}\label{eq:definition_Z(z;u)}
Z(z;u) \coloneqq X(z) + uY(z)
\end{align}

so that insertions of $Z(z;u)$\footnote{More accurately, insertions of operators which are $Q$-closed functions of $Z(z;u)$.} correspond holographically to boundary correlators with insertions at $z$ and $u$ of the respective $\bC P^1$.

From the considerations of the previous section, we know there are the $SL_L(2)$ and $SL_R(2)$ symmetries of the bulk that act separately on each $\bC P^1$. In the chiral algebra, these correspond respectively, to global conformal transformations and an $SL(2)$ R-symmetry mixing $X$ and $Y$ \cite{costello2021twistedholography}
\begin{align}\label{eq:R-symmetry}
{X \choose Y} \mapsto 
\begin{pmatrix}
\alpha & \beta\\
\gamma & \delta
\end{pmatrix}
{X \choose Y}; \qquad 
\begin{pmatrix}
\alpha & \beta\\
\gamma & \delta
\end{pmatrix} \in SL(2,\bC)
\end{align}

Expressed in terms of $Z(z;u)$, a global conformal transformations that maps $z \mapsto \frac{\delta z + \gamma}{\beta z +\alpha}$ acts on $Z$ as
\begin{align}
    Z(z;u) \mapsto \frac{1}{\beta z + \alpha} Z\left(\frac{\delta z + \gamma}{\beta z +\alpha};u\right)
\end{align}

Meanwhile, the R-symmetry \eqref{eq:R-symmetry} acts as
\begin{align}
    Z(z;u) \mapsto (\gamma u + \alpha)Z\left(z; \frac{\delta u + \beta}{\gamma u +\alpha}\right)
\end{align}

From these expressions, it is easy to see that these transformations act on $z$ and $u$ as the bulk $SL_L(2)$ and $SL_R(2)$ actions from equations \eqref{eq:SL2L_definition} and \eqref{eq:SL2R_definition}. Given this correspondence, we will henceforth refer to these chiral-algebra actions as their bulk counterparts $SL_L(2)$ and $SL_R(2)$.

As a final remark, we leave note of the OPEs of the $Z(z;u)$ fields
\begin{align}
Z^a_b (z_1;u_1)Z^c_d(z_2;u_2) \sim \frac{1}{N} \frac{u_1-u_2}{z_1-z_2}\, \delta^a_d \delta ^b_c  
\end{align}

% This intuition finds a nice holographic correspondence with the $SL_L(2)$, \eqref{eq:SL2L_definition} and $SL_R(2)$ \eqref{eq:SL2R_definition} actions of the previous section. More specifically, the $SL_2(\bC)$ of global conformal transformations of the chiral algebra with the $SL_L(2)$ action on the bulk, while the $SL_2(\bC)$ R-symmetry\footnote{This symmetry is a remnant of the physical theory's R-symmetry \cite{costello2021twistedholography}.} mixing $X$ and $Y$:
% \begin{align}
% {X \choose Y} \mapsto \begin{pmatrix}
% \alpha &\beta\\
% \gamma & \delta
% \end{pmatrix} {X \choose Y}; \quad \begin{pmatrix}
% \alpha &\beta\\
% \gamma & \delta
% \end{pmatrix}\in SL_2(\bC)
% \end{align}

% As $N$ goes to infinity, we'll take $\hbar$ to scale as  $1/N$ so that our OPE's have the appropriate $1/N$ scaling. \al{is there a more pedagogical or cleaner way to justify this? why is this the appropriate scaling?.}

\subsection{An easy way to build Q-closed operators}

The BRST transformation of our fields are:
\begin{align}
&Q X = [c,X] \label{eq_ad:QX}\\
&Q Y = [c,Y] \label{eq_ad:QY}\\
&Q c = \frac{1}{2}[c,c] = cc \label{eq_ad:Qc}\\
&Q b = [c,b] + [X,Y] \label{eq_ad:Qb}
\end{align}

which are generated by the BRST charge:
\begin{align}\label{eq_ad:BRST_charge}
Q &= N\oint  \dd z \,\Tr c[X,Y] + \frac{1}{2} \Tr b[c,c]
\end{align}

Given these transformations, one would naively expect that gauge invariant functions of $X$ and $Y$ would be an easy guess for $Q$-closed operators. Perhaps counter-intuitively, this expectation is wrong. To illustrate the point, let's take as an example the normal ordered operator $\mathcal{O}(z)=\Tr XYXY(z)$. We introduce momentarily a factor of $\hbar$ in propagators to keep track of wick contractions. The BRST action $[Q,\mathcal{O}]$ can be calculated knowing $Q$, \eqref{eq_ad:BRST_charge}, knowing the OPEs, \eqref{eq_ad:OPEs_U(N)}, and going through standard free 2d CFT calculations:
\begin{align}
[Q,\Tr XYXY(z)] &= \oint \frac{\dd z'}{2\pi i} \, \frac{1}{\hbar}\Tr c[X,Y](z')\, \Tr XYXY(z) \nonumber\\
&= \oint \frac{\dd z'}{2\pi i} \, \frac{1}{\hbar}\wick{\Tr c[X,\c1 Y](z')\, \Tr \c1 XYXY(z)} + \dots\text{(other one-wick-contraction terms)}+\nonumber\\
&\quad \oint \frac{\dd z'}{2\pi i} \, \frac{1}{\hbar}\wick{\Tr c[\c2 X,\c1 Y](z')\, \Tr \c1 X \c2 YXY(z)}+ \dots\text{(other two-wick-contraction terms)}\nonumber\\
&= 0 + 2 \hbar (\partial cYX - \partial cXY)
\end{align}

At tree level indeed we get that the action of $Q$ is 0. However, we see that quantum corrections in the form of a two-contraction term spoil the $Q$-closedness. This makes the problem of finding $Q$-closed operators somewhat non-trivial. There is however an obvious family of such operators: any gauge invariant function $f(X)$ of the $X$ field alone must be $Q$-closed since in

\begin{equation}
[Q,f(X)] \propto \int \dd z' \, \wick{\Tr c[X,\c1 Y](z')\, f(\c1 X)}\nonumber
\end{equation}
one cannot have more than one wick contraction.

Moreover, the $SL_R(2)$ symmetry \eqref{eq:R-symmetry} commutes with $Q$ so we can use it to rotate $X(z) \mapsto Z(z;u) = X(z)+uY(z) $ and $f(Z)$ will still be $Q$-closed. The determinant operators studied in this paper are of the class $f(Z(z;u))$.

% \subsection{Brief remarks on factors of \texorpdfstring{$\hbar$}{hbar}}\label{sec:brief_remarks_on_factors_of_hbar}

% Throughout this work, we introduce additional factors of $\hbar$ compared to the previous literature \cite{costello2021twistedholography},\cite{giant_gravitons_kasia}. In particular, not only OPEs but the argument of the determinant operators (Section \ref{sec:defining_determinants}) and baryons (Section \ref{sec:with_fundamentals_come_baryons}) will carry $\frac{1}{\hbar}$ factors
% \begin{align}
% \det\left(\frac{1}{\hbar}Z(z;u)\right) \qquad \varepsilon_{i_1\cdots i_N}\frac{J^{i_1}}{\hbar}\cdots\frac{J^{i_N}}{\hbar}
% \end{align}

% These factors, allow us to distinguish between the classical limit $\hbar \rightarrow 0$ and the 't Hooft limit $\hbar \rightarrow 0$ \emph{and} $N\rightarrow \infty$... \al{Elaborate}

% To the unwarned reader may seem an arbitrary cause of clutter

\section{A Complex for a Brane}\label{sec:A_Complex_for_an_Instanton}
In \cite{giant_gravitons_kasia}, the authors were able to match correlation functions of determinant operators to dual D1 branes. However, we believe that their construction has room for improvement in two aspects: First, the compatibility of their construction with the $SL_L(2)$ and $SL_R(2)$ symmetries of the problem could be made more manifest. Second, their construction for the dual D1 brane is not in terms of the natural object corresponding to branes in the B-model: Derived coherent sheaves. In this section we provide a construction that improves upon both these aspects.

Additionally, one may enrich the chiral algebra with degrees of freedom in the fundamental representation of the gauge group (see section \ref{sec:with_fundamentals_come_baryons}). Holographically, this corresponds to adding space-filling D5 branes in $SL_2(\bC)$. If both determinants \emph{and} fundamentals are present, one could wonder whether the D1-D5 brane system could admit novel deformations akin to those that lead to instantons in Dp-D(p-4)-brane systems \cite{Witten:1994tz_instantons_D_brane_systems}\cite{Douglas:1996uz_gauge_fields_and_d_branes}\cite{Douglas:1995bn_branes_with_branes}. In Section \ref{sec:the_instanton_complex} we propose one such family of deformations involving correlation functions of operators that are both determinant and baryonic in nature. To these correlation functions we assign an ADHM like complex, which we propose as our candidate dual D-brane system. 

Before showcasing the main constructions of this paper, we do a brief review of determinant operators and their correlation functions in the context of Twisted Holography.

% It is standard holographic lore that introducing fundamental degrees of freedom in your boundary theory corresponds to the presence of D-branes in the bulk. In CITE TWISTED HOLOGRAPHY, they studied 

% Holographically, the presence of flavours is expected to correspond to the presence of space-filling branes in the bulk. The brane created by standard determinant operators could simply be superimposed to these without any effect. 
 
% The combined set of branes, though, admits novel deformations akin to the ones which can deform a 
% D0-D4 system to a collection of instantons on the D4 brane world-volume. We expect these to map to 
% modifications of determinant operators involving the $I$ and $J$ fields. Our objective in this section is to 
% establish a dictionary \al{is it?} for such modifications and match the dual correlation functions in the planar approximation. 

\subsection{Review of Determinants and Saddles}\label{sec:review_of_determinants_and_saddles}
The basic determinant operators we make use of in this work take the form

\begin{equation}\label{eq:det_definition}
\det\left(m + Z(z;u) \right) \qquad\text{with} \qquad Z(z;u)=X(z) + u Y(z)
\end{equation}

Insertions of these correspond in the large N limit to the presence of probe D1-branes. This correspondence may be argued heuristically from a 't Hooft expansion perspective by writing the determinant as an integral:
\begin{align}\label{eq:fermionized_determinant}
	\int d \bar \psi d \psi e^{\bar \psi \left(m + Z \right)\psi}
\end{align}
with auxiliary fermionic degrees of freedom $\psi$ and $\psibar$ which, being in the fundamental and anti-fundamental representation of the gauge group, provide worldsheet boundaries in a large $N$ ribbon graph expansion.

This correspondence was made precise in \cite{giant_gravitons_kasia}, where they were able to map a correlation function of determinant operators to a vector bundle supported on a 2d surface in $SL_2(\bC)$ i.e. the corresponding dual brane. 
% In Section \ref{sec:a_brane_as_a_complex} we refine this proposal by mapping the correlation function to a vector bundle (more precisely a sheaf) defined cohomologically in terms of a complex of coherent sheaves \al{again, be wary of using this word}, a natural candidate for a brane in the B-model. Before doing so let's review the construction from \cite{giant_gravitons_kasia}.
The first step in their proposal uses techniques developed by \cite{Komatsu_2020} that allow to recast expectation values of determinant operators as integrals over auxiliary matrix degrees of freedom which are then amenable to a saddle point analysis. We strongly recommend \cite{Komatsu_2020}'s section 3 as a pedagogical introduction to the determinant operator calculations used in \cite{giant_gravitons_kasia} and in this work. Moreover, in Appendix \ref{app:detailed_SO_Sp_calculation} we provide a very detailed account of how these calculations proceed for the chiral algebras with $SO$ and $Sp$ gauge group.

To summarize briefly, the approach in \cite{Komatsu_2020} first expresses the determinants as fermionic integrals as in \eqref{eq:fermionized_determinant}, and subsequently integrates out the $X$ and $Y$ fields
\begin{align}
\left\langle\prod_\alpha :\det \left(m_\alpha + Z(z_\alpha;u_\alpha) \right): \right\rangle &= \left\langle \int d \bar \psi d \psi \,\prod_\alpha :e^{\bar \psi^\alpha \left(m_\alpha + Z(z_\alpha;u_\alpha) \right)\psi_\alpha }:\right\rangle\,\\
&= \int d \bar \psi d \psi e^{- \sum_{\alpha<\beta} \frac{1}{N}\frac{u_\alpha-u_\beta}{z_\alpha - z_\beta}\bar \psi^\beta\psi_\alpha \, \bar \psi^\alpha\psi_\beta}
\end{align}

Afterwards, they ``factorize" the quartic $\psi$ interaction as one mediated by a cubic interaction through an auxiliary field $\rho^\alpha_\beta$, through what is called  a Hubbard-Stratonovich transformation:
\begin{align}
\biggl\langle\prod_\alpha :&\det \left(m_\alpha + Z(z_\alpha;u_\alpha) \right): \biggr\rangle = \\
&\frac{1}{Z_\rho}\int d \bar \psi d \psi d \rho\,  e^{ \sum_{\alpha<\beta} N\frac{z_\alpha - z_\beta}{u_\alpha-u_\beta} \rho^\alpha_\beta \rho^\beta_\alpha + \sum_{\alpha\neq \beta} \rho^\alpha_\beta \bar \psi^\beta\psi_\alpha}
\end{align}
% \left[\prod_{\alpha < \beta} \frac{1}{2 \pi i } \frac{u_\alpha-u_\beta}{z_\alpha - z_\beta}\right]
where
\begin{align}
Z_\rho = \int d \rho\,  e^{ \sum_{\alpha<\beta} N\frac{z_\alpha - z_\beta}{u_\alpha-u_\beta} \rho^\alpha_\beta \rho^\beta_\alpha }
\end{align}

Finally, they integrate out the fermions:

\begin{align}\label{eq:integrate_out_psis}
\left\langle\prod_\alpha :\det \left(m_\alpha + Z(z_\alpha;u_\alpha) \right): \right\rangle = 
\frac{1}{Z_\rho}\int  d \rho \, e^{ \sum_{\alpha<\beta}N \frac{z_\alpha - z_\beta}{u_\alpha-u_\beta} \rho^\alpha_\beta \rho^\beta_\alpha + N \log \rho} 
\end{align}

Performing a saddle point analysis in the large $N$ limit, one finds this last integral is dominated by $\rho$'s which satisfy the saddle equation:

\begin{equation}\label{eq:U(N)_saddle_eq_component_form}
(z_\alpha - z_\beta) \rho^\alpha_\beta + (u_\alpha-u_\beta) (\rho^{-1})^\alpha_\beta = 0 
\end{equation}

which may be expressed in matrix form
\begin{equation}\label{eq:U(N)_saddle_eq_matrix_form}
[\zeta,\rho] + [\mu,\rho^{-1}] = 0
\end{equation}

in terms of the diagonal matrices: 
\begin{equation}\label{eq:saddle_eq}
\mu^\alpha_\beta = u_\alpha \delta^\alpha_\beta \qquad \text{and} \qquad \zeta^\alpha_\beta = z_\alpha \delta^\alpha_\beta
\end{equation}

We conclude with some remarks on how $\rho$ is acted by $SL_L(2)$ and $SL_R(2)$. There is some leeway on how one defines this transformation, as any action that preserves the saddle equation under the simultaneous change of \( z_\alpha \), \( u_\alpha \) and $\rho$, will do. We choose one of these and define the actions on $\rho$ to be
\begin{align}
\rho \xmapsto{SL_L(2)}(\beta \zeta +\alpha)\rho\,; \qquad \rho \xmapsto{SL_R(2)} \frac{1}{\gamma \mu + \alpha}\rho
\end{align}

\subsection{A Brane as a Spectral Curve}\label{sec:a_brane_as_a_curve}
\begin{figure}[h]
    \centering
    \includegraphics[width = 0.5\textwidth]{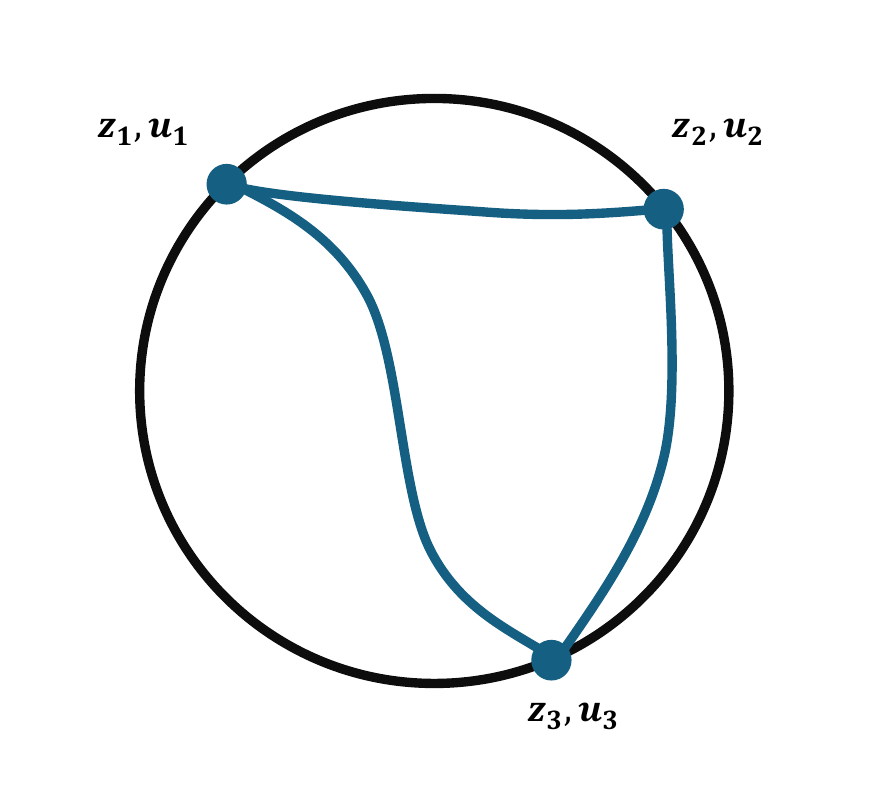}
    \caption{Schematic representation of the brane as a complex curve in $SL_2(\bC)$. The complex curve asymptotes to the boundary determinant insertions.}
    \label{fig:Dbrane}
\end{figure}

The holographic dictionary proposed by \cite{giant_gravitons_kasia} identifies the $\rho$ saddle with a complex curve in $SL_2(\bC)$. To do this, they built out of $\rho$ a set of commuting matrices, $B(a),C(a),D(a)$ whose eigenvalues span the complex curve as we vary $a$. To build these matrices they used the following facts:

\begin{enumerate}
\item The matrices should commute only when $\rho$ is a saddle, therefore, their commutator should be proportional to the saddle equation.

\item The complex curve should asymptote as $a$ goes to infinity, to the locations of the determinant insertions as shown schematically in figure \ref{fig:Dbrane}. This means the eigenvalues $b_\alpha(a)$ of $B(a)$ behave for large $a$ as (recall Eq \eqref{eq:boundary_CP1_coordinates} which relates the $u,z$ boundary coordinates to the bulk $a,b,c,d$ ones) $\frac{b_\alpha(a)}{a} = u_\alpha + O(\frac{1}{a})$, while those of $C(a)$ asymptote to $\frac{c_\alpha(a)}{a} = z_\alpha + O(\frac{1}{a})$.

\item The brane should live in $SL_2(\bC)$ which implies that the matrices should satisfy the contraint $aD(a) - B(a)C(a) = 1$. Essentially, this means one needs only find matrices $B(a)$ and $C(a)$ satisfying 1. and 2., and then define $D(a)\coloneqq \frac{1+B(a)C(a)}{a} = \frac{1+C(a)B(a)}{a}$.
\end{enumerate}

Expressing the equations from point 2. in matrix form and multiplying by $a$, we can conclude
\begin{align}
B(a) = a\mu + O(1)\\
C(a) = a\zeta + O(1)
\end{align}
Furthermore, if we denote the respective \( O(1) \) corrections as \( F_B(a) \) and \( F_C(a) \), point 1 imposes:
\begin{align}
[B(a),C(a)] &= a([\mu,F_C(a)] + [F_B(a), \zeta]) + [F_B,F_C] 
\\&\propto[\mu, \rho^{-1}] + [\zeta,\rho] 
\end{align}
for which an obvious solution is $F_B(a) = -\rho$ and $F_C(a) = \rho^{-1}$.
This gives the matrices
\begin{align}\label{eq:B,C,D}
B(a) &= \mu a - \rho \cr
C(a) &= \zeta a + \rho^{-1} \cr
D(a) &= \mu \zeta a - \zeta \rho +  \rho^{-1} \mu = \mu \zeta a - \rho \zeta +  \mu \rho^{-1} 
\end{align}

constructed in \cite{giant_gravitons_kasia} which satisfy all the desired properties. Note that no $1/a$ pole survived in defining $D(a)$, implying the matrix is well-defined over all of the curve, and hinting at the validity of the construction.

\subsection{A Brane as a Complex}\label{sec:a_brane_as_a_complex}

The previous presentation of the brane slightly obfuscates the $SL(2)_L \times SL(2)_R$ symmetries of the problem, simply because it gives the $a$ coordinate a preferential treatment. It is one of the main results of this work to provide a construction that respects these symmetries.

To do so let's rewrite the condition that the $B,C,D$ matrices have a common eigenvector as the statement that the set of matrices 
\begin{align}
B(a)-b &= \mu a -b - \rho \cr
C(a)-c &= \zeta a -c + \rho^{-1} \cr
D(a)-d &= \mu \zeta a -d - \zeta \rho + \rho^{-1} \mu 
\end{align}

have a common null vector. If we multiply the second matrix by $\mu$ on the right and subtract it from the third we get a more symmetric pair of matrices 
\begin{align}\label{eq:new_matrices}
M_1 &\coloneqq \mu a -b - \rho \cr
M_2 &\coloneqq \mu c -d - \zeta \rho 
\end{align}
which are co-variant under $SL(2)_L \times SL(2)_R$ and have a common null vector on the spectral curve. Alternatively, if we use the equivalent expression for \( D(a) = \mu \zeta a - \rho \zeta + \mu \rho^{-1} \) and multiply \( C - c \) by \( \mu \) from the left, we obtain the matrix 
\begin{align}
M_2' \coloneqq \mu c -d - \rho \zeta
\end{align}

We will use \( M_2' \) later in the construction of the complex associated with the brane. For now, observe that if \( M_1 \) and \( M_2 \) share a common null vector, then \( B \) and \( C \) must share common eigenvectors. This conclusion arises from the definition of \( M_1 \) for \( B \), and for \( C \), it follows from the relation:  
\begin{equation}	
cM_1 - aM_2  = c(\mu a - b - \rho ) - a(\mu c - d - \zeta \rho) = (\zeta a - c + \rho^{-1})\rho = (C(a) - c)\rho
\end{equation}
where in the equality we used the fact that $ad-bc = 1$.

We conclude that the two presentations of the spectral curve are fully equivalent, with the one in terms of $M_1$ and $M_2$, providing a more symmetric construction. 

Finally, a D-brane in the B-model is not uniquely characterized by its support: that information should be supplemented by a line bundle or vector bundle on the worldvolume. A candidate line bundle was guessed in \cite{giant_gravitons_kasia}, however, a better way to present a D-brane is as the cohomology of some chain complex. There is a way to do this starting from the spectral curve $B(a), C(a), D(s)$ matrices. Denote as $l$ the number of determinant insertions. We define the brane in terms of the complex  
\begin{equation}\label{eq:brane_complex}
\bC^{l} \xrightarrow{d_1(A)} \bC^{l} \oplus \bC^{l} \xrightarrow{d_2(A)} \bC^{l}
\end{equation}
with differentials given by
\begin{align}
d_1 = {B(a)-b\choose C(a)-c } ; \qquad
d_2 = \left(-C(a)+c \quad  B(a)-b\right) 
\end{align}
% where $A$ refers to the point in $SL_2(\bC)$
% \begin{equation*}
% A = \begin{pmatrix}
% a & b\\
% c & \frac{1+bc}{a}
% \end{pmatrix}  \in SL_1(\bC)
% \end{equation*}
In this way, at every point $A= \begin{pmatrix}
 a & b\\
 c & \frac{1+bc}{a}
\end{pmatrix}  $ in $SL_2(\bC)$ we can define the fiber of the Chan-Paton bundle as the cohomology $H_A \coloneqq \ker d_2(A) / \text{Im } d_1(A)$. Note that the differentials square to zero precisely when $B$ and $C$ commute, $d_2 d_1 = [B(a), C(a)]$. Furthermore, the cohomology is non-trivial only when the kernel of $d_2$ is non-trivial, which happens only when $b$ and $c$ are eigenvalues of $B(a)$ and $C(a)$ respectively. Therefore, our Chan-Paton bundle is non-trivial only over the spectral curve of \cite{giant_gravitons_kasia}.

Furthermore, just as $M_1$, $M_2$ and $M_2'$ provide a more symmetric presentation of the spectral curve, they also may be used to construct a complex that preserves the $SL_L(2)$ and $SL_R(2)$ symmetries of the problem. The new differentials are
\begin{align}
d_1 = {M_1\choose  M_2'} ; \qquad
d_2 = \left(M_2\rho^{-1}  \quad  -M_1\rho^{-1} \right) 
\end{align}

which can be written explicitly as
\begin{align}%\label{eq:brane_differentials}
d_1 &= {\mu a - b - \rho \choose \,\,\mu c -d - \rho\zeta} \label{eq:brane_d1}\\
d_2 &= \left((\mu c - d)\rho^{-1} - \zeta \quad -(\mu a- b)\rho^{-1} + 1\right)\label{eq:brane_d2}
\end{align}

The condition $d_2 d_1=0$ is guaranteed by the fact that $\rho$ is a saddle 
\begin{align}\label{eq:d2d1=0}
d_2 d_1 = -M_1\rho^{-1}M_2' + M_2 \rho^{-1} M_1 = [\zeta, \rho]+[\mu, \rho^{-1}]
\end{align}

We propose this complex as a very natural candidate to define the D1 brane dual to a given saddle, which additionally transforms covariantly under the $SL_L(2)$ 
\begin{align}
d_1 \mapsto 
\begin{pmatrix}
\alpha & \beta \\
\gamma & \delta
\end{pmatrix}
d_1; \qquad
d_2 \mapsto \frac{1}{\beta \zeta + \alpha} d_2 
\begin{pmatrix}
\alpha & \beta \\
\gamma & \delta
\end{pmatrix}^{-1}
\end{align}
and $SL_R(2)$
\begin{align}
d_1 \mapsto d_1\frac{1}{\gamma \mu + \alpha} ; \qquad 
d_2 \mapsto d_2
\end{align}

\subsection{With Fundamentals come Baryons}\label{sec:with_fundamentals_come_baryons}

Our chiral algebras admit interesting modifications where we add fundamental fields $J$ and anti-fundamentals $I$. % with action:% \begin{align}
% \int \dd^2z \, I\bar\partial J
% \end{align}
The classical BRST transformations act on $I$ and $J$ by gauge transformations with parameter $c$. This translates into an extra term in the BRST charge $\oint\Tr IcJ$ which may spoil nilpotency of $Q_{\text{BRST}}$ quantum mechanically. 
In order to cancel such anomalies, we need to add both bosonic and fermionic fields in equal number \cite{costello2021twistedholography}. We denote the number of fundamental fields of each type as $F$ so that the resulting chiral algebra classically has $U(F|F)$ global symmetry. When necessary, we can denote individual components as $J^a_A$ and $I^b_j$ with $A,B$ the $U(F|F)$ flavor indices and  $a,b$ gauge indices.
It is a bit tedious to keep track of signs from Grassmann parity, so to lighten expressions we will treat the $I$ and $J$ fields as if they were Grassmann odd. 

Their OPE is given by  
\begin{equation}
J^a_A(z) I^B_b(w) \sim \frac{\delta^a_b \delta^B_A}{z-w}
\end{equation}

% The BRST current now has an extra $\Tr c J I$ term. The tree-level BRST charge acts on $I$ and $J$ by gauge transformations with parameter $c$. 

% The currents for the $U(F|F)$ symmetry are just $j^A_B \equiv I^A J_B$. They are $Q$-closed in the $SU(N)$ theory. In a $U(N)$ theory they would have an anomaly forcing them to be traceless and the diagonal current would be $Q$-exact. 

% This sort of anomaly affects the mesonic operators 
% \begin{equation}
% 	E^A_B(u;z) \equiv  I^A (X + u Y)^n J_B
% \end{equation}
% These are $Q$-closed at tree level, but at 1-loop we can contract $I$ and $J$ to get something like $\delta^A_B \Tr \partial c (X + u Y)^n$. Furthermore,
% $Q_0 \Tr b (X + u Y)^n = N^{-1} E^A_A(u;z)$. Hence only the traceless part of the mesons is $Q$-closed and the meson proportional to $\delta^A_B$ can be rewritten in terms of traces \al{To do: elaborate further upon this so I can see more crisply which operators are Q-closed and which fail to be}. 

% The action of $Q$ on operators built from $X$ and $Y$ is unaffected by the extra flavours, so that correlation functions of 
% $A_n$'s or determinant operators are unchanged. 

% There is a great deal of holographic information that can be extracted from these \al{COMPLETE} 
Holographically, the presence of flavours is expected to correspond to
the presence of space-filling branes in the bulk \cite{costello2021twistedholography}. The brane created by standard determinant operators could simply be superimposed to these 
 without any effect. 
 
The combined set of branes, though, admits novel deformations akin to the ones which can deform a 
D0-D4 system to a collection of instantons on the D4 brane world-volume \cite{Witten:1994tz_instantons_D_brane_systems}\cite{Douglas:1996uz_gauge_fields_and_d_branes}\cite{Douglas:1995bn_branes_with_branes}. We expect these to map to 
modifications of determinant operators involving the $I$ and $J$ fields. Our objective in this section is to establish a dictionary between such modifications and a candidate bulk brane system. 

% Modifications mixing both branes, should involve both the fundamental dofs introduced by the determinants $\bar\psi, \psi$ and the fundamental flavors $I,J$. Obvious deformation candidates, are deformations to the action of our theory by the terms $\bar\psi J(z) \epsilon$ and $\bar\epsilon I(z)\psi$, with $z$ the location of the determinant insertion, and $\epsilon$ and $\bar\epsilon$ some couplings. \al{improve}. These correspond to insertion of baryonic 
% % These modifications will be a mixture of the determinant operators of the previous section and baryonic operators such as 
% \begin{equation}
% \label{eq:baryonic}
% \int [d\bar\psi] e^{\psi J \epsilon} = \varepsilon_{i_1 \cdots i_N} J^{i_1}_{A_1}\epsilon^{A_1} \cdots J^{i_n}_{A_n}\epsilon^{A_n}
% \end{equation}
% and correspondingly antibaryonic operators
% \begin{equation}
% \label{eq:anti-baryonic}
% \int [d\bar\psi] e^{ \bar\epsilon I \psi}
% \end{equation}

% However, for these to be $Q$-closed, we'll need for the gauge group of our chiral algebra to be  $SU(N)$ and not $U(N)$. A few remarks about this new chiral algebra are then in order.

Modifications that mix both branes should involve the fundamental degrees of freedom introduced by the determinants, \( \bar\psi \) and \( \psi \), as well as the fundamental flavors \( I \) and \( J \). Natural candidates for such deformations include terms such as \( \bar\psi J(z) \epsilon \) and \( \bar\epsilon I(z) \psi \), where \( z \) denotes the location of the determinant insertion and \( \epsilon \), \( \bar\epsilon \) are coupling parameters. These correspond to the insertion of baryonic operators
\begin{equation}
\label{eq:baryonic}
\int [d\bar\psi] e^{\psi J \epsilon} = \varepsilon_{i_1 \cdots i_N} J^{i_1}_{A_1}\epsilon^{A_1} \cdots J^{i_n}_{A_n}\epsilon^{A_n}
\end{equation}
and the corresponding antibaryonic ones
\begin{equation}
\label{eq:anti-baryonic}
\int [d\bar\psi] e^{ \bar\epsilon I \psi}
\end{equation}
For these operators to be \( Q \)-closed, the gauge group of the chiral algebra must be \( SU(N) \) rather than \( U(N) \). Consequently, a few remarks about this new chiral algebra are in order.  

The new OPE in the $SU(N)$ theory is modified to 
\begin{equation}
Y^a_b(z) X^c_d(w) \sim \frac{1}{N}\left[\delta^a_d \delta^c_b-\frac{\delta^a_b \delta^c_d}{N}\right]\frac{1}{z-w}
\end{equation}
Gauge invariant functions of $Z(z;u)=X(z)+uY(z)$ are still $Q$-closed as the reasoning used for $U(N)$ still holds, so determinant operators of $Z$ are still in cohomology.  However, their correlation functions acquire subleading corrections
\begin{align}
\Biggl<\prod_i &\det \left(m_i + Z(z_i;u_i)\right) \Biggr> = \cr &= \int d \bar \psi d \psi e^{- N\sum_{i<j} \frac{u_i-u_j}{z_i - z_j}\bar \psi_j\psi^i \, \bar \psi_i\psi^j-\frac{1}{N^2} \sum_{i<j} \frac{u_i-u_j}{z_i - z_j}\bar \psi_j\psi^j \, \bar \psi_i\psi^i + \sum_i m_i\psibar_i\psi^i }%:\,e^{\sum_i \bar \psi^i \left(m_i + X(z_i) + u_i Y(z_i) \right)\psi_i }\,:
\end{align}
We can account for them by acting with  
\begin{equation}
	e^{-\frac{1}{N^2} \sum_{i<j} \frac{u_i-u_j}{z_i - z_j}\partial_{m_i} \partial_{m_j}}
\end{equation}
but this does not affect the semi-classical saddle and contributes only to subleading contributions beyond the scope of this work.

% Returning to the baryons, we can express them as fermionic integrals
% \begin{equation}
% \varepsilon_{i_1 \cdots i_N} J^{i_1}_{A_1}\epsilon^{A_1} \cdots J^{i_n}_{A_n}\epsilon^{A_n} = \int d \bar \psi e^{\bar \psi J \epsilon}
% \end{equation}
% where $\epsilon^A$ are parameters. We can similarly fermionize an anti-baryon:
% \begin{equation}
% 	\int d\psi e^{\bar \epsilon I\psi}
% \end{equation}
% with parameters $\bar \epsilon_A$. 

% For example, the correlation function of a baryon and an anti-baryon is 
% \begin{equation}
% \left\langle \int d \bar \psi d\psi  e^{\bar \psi J(z) \epsilon+\bar \epsilon I(w) \psi}\right\rangle =\int d \bar \psi d\psi  \,e^{\frac{1}{z-w}  \bar \epsilon \epsilon \bar \psi \psi} 
% \end{equation}

Returning to our previous discussion, we can try to construct a local operator that combines a baryon and an anti-baryon: 
\begin{equation}
\int d \bar \psi d \psi e^{\bar \psi J \epsilon+\bar \epsilon I \psi}
\end{equation}
This operator is no longer $Q$-closed however; there is a 1-loop anomaly proportional to $\bar \epsilon \epsilon$, that we can nevertheless cancel by requiring $\bar \epsilon \epsilon =0$.

% Such an operator can be expanded into a polynomial in $IJ$ currents, in the same way as a determinant can be expanded into traces \al{don't say this if you're not saying it about determinants}. 
% We can also deform it to
% \begin{equation}
% \int d \bar \psi d \psi e^{m \bar \psi \psi + \bar \psi J \epsilon+\bar \epsilon I \psi} = m^N e^{\frac{\bar \epsilon IJ \epsilon}{m}}
% \end{equation}
More interestingly, we can combine these with the determinant operators
\begin{equation}\label{eq:det_baryon_operator}
\mathcal{O}(m,z,u,\epsilon,\bar\epsilon) \coloneqq\int d \bar \psi d \psi e^{\bar \psi \left(m +Z(z;u) \right)\psi+ \bar \psi J \epsilon+\bar \epsilon I \psi}
\end{equation}
so that we have a deformation that interpolates between a determinant and a product of a baryon and an anti-baryon. We need again  $\bar \epsilon \epsilon =0$ in order to avoid an anomaly. These are natural candidates to describe a deformation of the bulk brane system. 

\subsection{A Complex for a D-Brane System}\label{sec:the_instanton_complex}

The expectation value of normal ordered products of these generalized determinant operators \eqref{eq:det_baryon_operator} can be computed as before. The contractions between $I$'s and $J$'s will now contribute the extra terms: 
\begin{equation}
	\bra\prod_\alpha \mathcal{O}_\alpha\ket = \int[\dd\psibar][\dd\psi]  e^{\cdots \,+ \sum_{\alpha\neq \beta} \frac{1}{z_\alpha-z_\beta}    \bar\epsilon_\beta\epsilon^\alpha \psibar_\alpha  \psi^\beta }
\end{equation}
one then may perform the same Hubbard-Stratonovich transformation and integrate out the $\psibar \psi$'s

\begin{equation}
\bra\prod_\alpha \mathcal{O}_\alpha\ket = \frac{1}{Z_\rho}\int [\dd\rho ]\exp\left(\frac{N}{2}\sum_{\alpha\neq \beta}\frac{z_\alpha - z_\beta}{u_\alpha - u_\beta}\rho^\alpha_\beta\rho^\beta_\alpha + N\Tr\left(\ln \left(\rho + \frac{1}{z-z}\epsilon\bar\epsilon\right)\right)\right)\\
\end{equation}
which after a shift $\rho^\alpha_\beta \mapsto \rho^\alpha_\beta - \frac{1}{z_\alpha - z_\beta} \epsilon^\alpha\bar\epsilon_\beta$, leads to a modified saddle equation:
\begin{equation}
(z_\alpha - z_\beta)\rho^\alpha_\beta + (u_\alpha- u_\beta)(\rho^{-1})^\alpha_\beta  = \epsilon^\alpha\bar\epsilon_\beta 
\end{equation}

which can be expressed in matrix form
\begin{align}
[\zeta,\rho] + [\mu, \rho^{-1}] = \epsilon\bar\epsilon
 \label{eq:SEflavour}
\end{align}

with $\zeta$ and $\mu$ the same diagonal matrices with eigenvalues $z_\alpha$ and $u_\alpha$ respectively. Note that the consistency of the equations for $\alpha=\beta$ only holds if the BRST condition $\epsilon^\alpha\bar\epsilon_\alpha = 0$ (with no summation over $\alpha$) is in place.

If we write as before 
\begin{align}
B(a) &= a \mu -  \rho\\
C(a) &= a \zeta + \rho^{-1} 
\end{align}
then $B$ and $C$ no longer commute. Instead, they satisfy 
\begin{equation}\label{eq:F-term_constraint}
[B(a), C(a)] = a \epsilon\bar\epsilon
\end{equation}
which resembles the F-term constraints on the data of an ADHM construction of an instanton configuration. The corresponding ADHM  construction would use a complex of the form 
\begin{equation}
	\bC^l \xrightarrow{d_1} \bC^{l} \oplus \bC^{l} \oplus \bC^{F} \xrightarrow{d_2} \bC^l
\end{equation}
with maps $d_1$ and $d_2$
\begin{align}
d_1 = \begin{pmatrix}
B(a)-b \\
C(a)-c \\
a\bar\epsilon
\end{pmatrix} ; \qquad
d_2 = \left(-C(a)+c \quad  B(a)-b \quad -\epsilon\right) 
\end{align}
satisfying $d_2d_1 = 0$ thanks to the F-term constraint \eqref{eq:F-term_constraint}.
We find that there is also a natural extension of the $SL_L(2)$ and $SL_R(2)$ covariant differentials, \eqref{eq:brane_d1} \eqref{eq:brane_d2}, that incorporates the new deformation:
\begin{align}\label{eq:instanton_complex}
d_1 &= 
\begin{pmatrix}
-\rho^{-1}(\mu c - d) + \zeta \\
\,\,\,\,\,\rho^{-1}(\mu a -b) - 1\\
-\bar\epsilon
\end{pmatrix}\\
d_2 &= \left(\mu a - b -\rho \quad \mu c -d -\zeta\rho \quad \epsilon\right)
\end{align}
We therefore propose this as our candidate D-brane system holographic dual to a specific saddle for the correlation function of generalized determinants.

\section{Twisted Holography with \texorpdfstring{$SO$ and $Sp$}{SO and Sp} gauge group}\label{sec:SO_Sp}
Now we proceed to apply the tools developed in the previous section to the chiral algebras with gauge group $SO(N)$ and $Sp(2N)$. We focus on the $Sp$ case for the sake of narrative, since the construction for the $SO$ case is completely analogous. 

To get intuitions about their holographic duals, we may look momentarily at their untwisted counterparts. When the scalars are in the adjoint representation of $Sp(2N)$, the chiral algebra arises from the twist of $\mathcal{N} = 4$ SYM with gauge group $Sp(2N)$. This last theory is holographically dual to an orientifold of Type IIB String Theory in $AdS_5\times\bR\bP^5$ \cite{Witten_1998_baryons_and_branes}. This leads to the conjecture of \cite{Jake_Abajian:2024rjq} that the holographic dual to these $Sp(2N)$ chiral algebras with adjoint matter is an orientifold of Kodaira Spencer theory with background $SL_2(\bC)/\bZ_2$.

% Instead, when $X$ and $Y$ live in the \emph{antisymmetric} representation, we find a different holographic dual. This chiral algebra arises from the twist of a superconformal $\mathcal{N}=2$ supersymmetric $Sp(2N)$ gauge theory with a hypermultiplet in the antisymmetric representation of $Sp(2N)$ and eight hypermultiplets in the fundamental representation of the gauge group. The holographic dual to this theory is Type IIB on a different orientifold of $AdS_5\times S^5$ where on reflects under the $\bZ_2$ action only a subset of the directions orthogonal to the brane. This leads to the conjecture that the holographic dual to the $Sp(2N)$ chiral algebras with antisymmetric matter should be the Type I topological string \cite{costello2020anomalycancellationtopologicalstring} (that is, the twist of Type I String Theory) in $SL_2\bC$.

Instead, when $X$ and $Y$ live in the \emph{antisymmetric} representation, we find a different holographic dual. The chiral algebra arises from the twist of an $\mathcal{N}=2$ theory with antisymmetric matter (plus some additional matter to make the theory super-conformal). In the bulk, one has an orientifold where only two of the six directions orthogonal to the brane are flipped. The twist of Type IIB happens to localize entirely on the fixed locus of this reflection, meaning the twisted theory doesn't see any $\bZ_2$ reflections of the orientifold, only the gauging of parity. This leads to the conjecture of \cite{Jake_Abajian:2024rjq} that the holographic dual to this chiral algebra is the Type I String \cite{costello2020anomalycancellationtopologicalstring} on $SL_2(\bC)$.   

% . However, one may find a conjecture for the holographid dual  \cite{costello2020anomalycancellationtopologicalstring} in $SL_2\bC$. \al{Should i elaborate?}

Note that the orientifold of Kodaira Spencer on $SL_2\bC/\bZ_2$ is not the same theory as the Type I Topological String, in particular, the Type I theory contains fewer fields. This is due to the fact that the Type I strings arises from gauging parity in the worldsheet B-model. In this way, one must drop from the spectrum all the fields that are charged under this $\bZ_2$ action. In contrast, no field is dropped in the Kodaira Spencer orientifold, instead, their values at points related by the orbifold $\bZ_2$ action are identified.

Our brane constructions provide evidence for the previous conjectures. However, before describing them we make some introductory remarks about these chiral algebras with gauge group $Sp(2N)$.

\subsection{Chiral Algebras with \texorpdfstring{$SO$ and $Sp$}{SO and Sp} Gauge Group}

% The main difference in going from $U(N)$ to $SO(N)/Sp(2N)$ is the change in how indices are paired in the OPEs

% \begin{align}
% Y^{ab}(z) X^{cd}(z') \sim
% \begin{cases}
%     \dfrac{\hbar}{z - z'}\;\dfrac{\delta^{ac}\delta^{bd} \pm \delta^{ad}\delta^{bc}}{2} & \text{for $SO(N)$}\\[9pt]
%     \dfrac{\hbar}{z - z'}\; \dfrac{\Omega^{ac}\Omega^{bd} \pm \Omega^{ad}\Omega^{bc}}{2} & \text{for $Sp(2N)$}
%     \end{cases}
% \end{align}

% with the $\pm$ depending on whether $X^{ab}$ and $Y^{ab}$ are symmetric, $X^{ab}=X^{ba}$, or anti-symmetric, $X^{ab}=-X^{ba}$, matrices.

% $b$ and $c$ see a similar modification of their OPEs:

% \begin{align}\label{eq_ad:OPEs}
%     b^{ab}(z) c^{cd}(z') \sim
% \begin{cases}
%     \dfrac{\hbar}{z - z'}\;\dfrac{\delta^{ac}\delta^{bd} - \delta^{ad}\delta^{bc}}{2} & \text{for $SO(N)$}\\[9pt]
%     \dfrac{\hbar}{z - z'}\; \dfrac{\Omega^{ac}\Omega^{bd} + \Omega^{ad}\Omega^{bc}}{2} & \text{for $Sp(2N)$}
%     \end{cases}
% \end{align}

% where we have no $\pm$ freedom since they must live in the adjoint representation of the respective gauge group.

Here we review some of properties of chiral algebras with gauge group $SO(N)$ and $Sp(2N)$, where again we focus on the $Sp(2N)$ chiral algebra to avoid redundancy. Having fixed the gauge group, we describe two possible representations for the $X$ and $Y$ fields: adjoint (aka symmetric) and antisymmetric.

The main difference in these new chiral algebras with respect to the $U(N)$ case is how indices pair up
\begin{align}
Y^{ab}(z) X^{cd}(z') \sim \frac{1}{N}\dfrac{1}{z - z'}\;\dfrac{\Omega^{ac}\Omega^{bd} \pm \Omega^{ad}\Omega^{bc}}{2} 
\end{align}
with the $\pm$ depending on whether $X^{ab}$ and $Y^{ab}$ are symmetric, $X^{ab}=X^{ba}$, or anti-symmetric, $X^{ab}=-X^{ba}$.

The expressions of the BRST transformations are identical to those of the $U(N)$ case and so is the BRST charge
\begin{align}
Q &= N\oint  \dd z \,\Tr c[X,Y] + \frac{1}{2} \Tr b[c,c]
\end{align}

If $X,Y$ are in the adjoint representation, the BRST charge is nilpotent at the quantum level. The same cannot be said for their antisymmetric counterparts; quantum mechanically there is a BRST anomaly
\begin{align}\label{eq:BRST_anomaly}
Q^2 =  -4\oint \Tr(\partial c c)
\end{align}
Not all is lost for these theories, however. The situation can be ameliorated by introducing additional dofs whose own BRST anomaly cancels the one arising from the chiral algebra. In particular, we can add fundamentals $I^a_A$, where $a$ is a gauge index and $A$ a flavor index, with OPE
\begin{align}\label{eq:II_OPE}
I^a_A(z) I^a_B(z') \sim \dfrac{1}{z - z'}\;\Omega^{ab}\eta_{AB}  
\end{align}
Consistency of the OPE demands that the bilinear pairing $\eta_{AB}$ must be symmetric (antisymmetric) for $I$ bosonic (resp. fermionic).

To cancel the anomaly \eqref{eq:BRST_anomaly} we introduce 8 of these bosonic flavors. Furthermore, we may introduce additional flavors in pairs, one bosonic and one fermionic, whose anomalies mutually cancel as in the $U(N)$ case. All and all, we end up with an $OSp(k+8|k)$ flavor symmetry, with $k$ the number of extra flavor pairs introduced beyond the 8 bosonic ones necessary to cancel the anomaly coming from the chiral algebra.

This business of anomaly cancellation finds a neat correspondence in the bulk. As already alluded, for chiral algebras with fields $X,Y$ in the antisymmetric representation of $Sp(N)$, we expect the holographic dual to be the Type I topological string defined in \cite{costello2020anomalycancellationtopologicalstring}. However, for this theory to be quantized consistently, one must introduce 8 space-filling branes, that is, couple the Type I theory to Holomorphic-Chern Simons with gauge group $SO(8)$. These branes are holographically dual to the eight flavors we must add in the boundary chiral algebra.

Moreover, corresponding anomaly cancellation mechanisms can be identified in the untwisted theories. In the physical \(\mathcal{N}=2\) theory, a sufficient number of hypermultiplets must be introduced to cancel the beta function and preserve conformal symmetry. Similarly, in the bulk, the specific orientifold of Type IIB introduces an \(O7\)-plane that generates an RR flux, which must be canceled by introducing eight \(D7\)-branes. This stack has an $SO(8)$ gauge symmetry, which becomes the $SO(8)$ gauge group in the Type I string after the twist.  

A similar story occurs in the chiral algebras with gauge group $SO(N)$; when $X,Y$ are in the adjoint, the theory is anomaly-free. In contrast, when they are in the antisymmetric representation, one must also introduce 8 additional \textit{fermionic} flavors to have a nilpotent BRST charge. In this case, however, the untwisted theory is not well studied as the associated ``physical" \(\mathcal{N}=2\) theory is non-unitary, requiring hypermultiplets that violate spin-statistics to maintain conformal symmetry. It is mainly due to this fact that our presentation focuses on the $Sp(2N)$ chiral algebra.

% This fact doesn't have a neat correspondence in the physical theories simply because one would have to introduce field content that violates spin-statistics and therefore unitarity. 

\subsection{Determinants and Baryons for \texorpdfstring{$SO$ and $Sp$}{SO and Sp} } \label{subsec:dets_SO_Sp}

Regarding $Q$-closed operators, gauge invariant functions $f(Z(z;u))$ still pass the test. Determinants become:
\begin{align}
\det(Z) = \int [d\psi d\psibar] \exp(\psibar_a Z^a_b \psi^b )
\end{align}
Focusing momentarily on the case with $Z^{ab}$ symmetric, we can use  our capacity to raise and lower $a,b$ indices with the symplectic form to re-express the right-hand side in the suggestive manner
\begin{align}
\int [d\psi] \exp\left(\frac{1}{2}\psi_{ai} Z^{ab}\omega^{ij} \psi_{bj} \right)
\end{align}
with  $\psi_{a1} = \psi_a$ and $\psi_{a2} = \psibar_a$, and with $\omega^{ij}$ the $2\times2$ obvious antisymmetric matrix. Having introduced the new $i,j$ indices, we might as well let them run from $2$ to $2k$. This means the determinant branes became dual to a stack with $2k$ branes. The anti-symmetric form $\omega^{ij}$, tells us the Chan-Paton bundle on this brane has gauge group $Sp(2k)$. Conversely, if $Z^{ab}$ is antisymmetric, the bilinear form pairing $i,j$ indices must be symmetric. We set it to be $\delta^{ij}$ w.l.o.g, in which case the determinant operators become
\begin{align}
\int [d\psi] \exp\left(\frac{1}{2}\psi_{ai} Z^{ab}\delta^{ij} \psi_{bj} \right)
\end{align}
We write both cases as 
\begin{align}
\int [d\psi] \exp\left(\frac{1}{2}\psi_{ai} Z^{ab} \tsr{\psi}{_b^i} \right)
\end{align}
where one must be careful to keep track of which bilinear form one is using to lower and raise the $i,j$ indices.

The mass term in the determinants of the $U(N)$ case (Eq. \eqref{eq:det_definition}) now becomes a matrix
\begin{align}
\int [d\psi] \exp\left(\frac{1}{2}\psi_{ai}m^{ij}\Omega^{ab}\psi_{bj} + \frac{1}{2}\psi_{ai} Z^{ab}\tsr{\psi}{_{b}^i} \right)
\end{align}
which is symmetric, as the antisymmetric part decouples.

Finally, we can combine the determinant with the $I$ fields to form our determinant-baryon mixture
\begin{align}
\int [d\psi] \exp\left(\frac{1}{2}\psi_{ai}m^{ij}\Omega^{ab}\psi_{bj} + \frac{1}{2}\psi_{ai} Z^{ab}\tsr{\psi}{_{b}^i} + \tsr{\psi}{^i_a} I^a_A \tsr{\epsilon}{^A_i}\right)
\end{align}
% \begin{align}
% \int d\psi e^{\psi \left(m + Z(z;u) \right)\psi+ \psi I \epsilon}
% \end{align}

In this case, the operator is $Q$-closed as long as we impose
\begin{align}
\eta_{AB}\tsr{\epsilon}{^A_i}\tsr{\epsilon}{^B_j}=0
\end{align}
with $\eta_{AB}$ the pairing of the $II$ OPE \eqref{eq:II_OPE}.
% \begin{equation}
% \epsilon_A\epsilon_B \delta^{AB}=0
% \end{equation}

\subsection{A Complex for an \texorpdfstring{$SO/Sp$}{SO/Sp} Brane System}

We proceed to construct the complex for the brane system dual to the expectation value of generalized determinant operators. We again focus on the gauge group $Sp(2N)$, the construction for $SO$ being completely analogous.

In both the cases where $X$, $Y$ are symmetric or antisymmetric the calculation of the expectation values proceeds almost identically to the $U(N)$ case and is shown in detail in Appendix \ref{app:detailed_SO_Sp_calculation}. When the dust settles, correlation functions of determinant/baryon operators can be recast into the by-now-familiar form
\begin{align}
\left\langle\prod_\alpha \right.&\left.\int [d\psi] :e^{\psi_\alpha \left(m_\alpha + Z(z_\alpha;u_\alpha) \right)\psi_\alpha+ \psi_\alpha I(z_\alpha) \epsilon_\alpha}:\right\rangle = \\
&\frac{1}{Z_\rho}\int [\dd\rho ]\exp\left(\frac{N}{4}\sum_{\alpha\neq \beta}\frac{z_\alpha - z_\beta}{u_\alpha - u_\beta}\rho_{\alpha i\beta j}\rho^{\beta j\alpha i} + \frac{N}{2}\Tr\left(\ln \left(\rho_{i\alpha j\beta} + \frac{1}{z_\alpha-z_\beta}\epsilon_{i\alpha A}\tsr{\epsilon}{_j_\beta^{A}}\right)\right)\right)
\end{align}
where we raise and lower $\alpha,\beta$ indices with the pairing $\delta^{\alpha\beta}$. 

This integral is dominated by the saddle
\begin{equation}
(z_\alpha - z_\beta)\rho_{\alpha i\beta j} + (u_\alpha- u_\beta)(\rho^{-1})_{\alpha i\beta j}  = \epsilon_{\alpha i A}\epsilon_{\beta j B}\Omega^{AB} %\quad \text{equivalently} \quad [\zeta,\rho] + [\mu, \rho^{-1}] = \epsilon \Omega^{-1}\epsilon^T
 \label{eq:saddle_equation_SO_Sp}
\end{equation}
which can be expressed in matrix form as
\begin{align}
[\zeta, \rho] + [\mu, \rho] = \epsilon \Omega^{-1} \epsilon^T 
\end{align}
with 
\begin{align}
\tsr{\zeta}{^{i\alpha}_{j\beta}} &= z_\alpha \delta^i_j \delta^\alpha_\beta\\
\tsr{\mu}{^{i\alpha}_{j\beta}} &= u_\alpha \delta^i_j \delta^\alpha_\beta
\end{align}
The main difference w.r.t. the $U(N)$ case is that now $\rho$ has symmetry properties: 
\begin{align}
\rho_{\beta j\alpha i} = \rho_{\alpha i \beta j} 
\end{align}

If we raise the $i$ and $\alpha$ indices, this expression can be written in matrix notation as 
\begin{align}
-\omega^{-1} \rho^T \omega &= \rho \qquad \text{for $X,Y$ symmetric} \label{eq:rho_syms_1}\\
\rho^T &= \rho \qquad  \text{for $X,Y$ antisymmetric}\label{eq:rho_syms_2}
\end{align} 
% $$, for $X,Y$ symmetric, and $\rho^T = \rho$, for $X,Y$ antisymmetric.

% when $X,Y$ are in the adjoint of $Sp(2N)$, $\rho$ satisfies, $\rho^T = -\omega\rho\omega^{-1}$ \al{elaborate a bit here} \al{also, what bilinear form did I use to raise the $\alpha$ and $\beta$ indices???}. Otherwise, when $X,Y$ are antisymmetric matrices, $\rho$ is symmetric, $\rho^T = \rho$.

% Note that $\omega$ commutes with $\zeta$ and $\mu$ since they act non-trivially on different indices. This point is worth keeping in mind to prove that the diagrams defined below commute.

Note that \(\omega\) commutes with \(\zeta\) and \(\mu\) as they act on different indices. This is a helpful detail to remember when verifying that the diagrams defined below \ref{eq:commuting_diagrams} commute.

As discussed in the previous section, it is conjectured in \cite{Jake_Abajian:2024rjq} that each chiral algebra corresponds to a distinct holographic dual: The adjoint one is dual to an orientifold of the B-model on $SL_2(\bC)/\bZ_2$, while the antisymmetric one is dual to the Topological Type I string in $SL_2(\bC)$. The relevant distinction between the two is that the latter gauges parity, the former gauges parity \textit{and} orbifolds. Gauging parity means the Chan-Paton bundle should be identified with its dual. The additional orbifold implies this identification happens over points related by the orbifold $\bZ_2$ action. Now we show how the corresponding brane complexes have precisely these identifications. 

Define the spaces $V^{-2} \coloneqq \bC^{l\times k} \eqqcolon V^{0}$ and $V^{-1} \coloneqq \bC^{l\times k}\oplus \bC^{l\times k}\oplus\bC^{F}$, where $l$ is the number of determinant insertions, $k$ is the size of the stack of the probe branes (i.e. the range of the $i,j$ indices), and $F$ the number of flavors. Then, the instanton complex for both reads:
\begin{equation}
V^{-2} \xrightarrow{d_{-1}(A)} V^{-1} \xrightarrow{d_{0}(A)} V^{0}
\end{equation}
where $A$ refers to the pointis the matrix in $SL_2(\bC)$ the differentials are evaluated at. 
The differentials for the adjoint chiral algebra, are given by:
\begin{align}
d_{-1}(A) &= 
\begin{pmatrix}
-\rho^{-1}(\mu c - d) + \zeta \\
\,\,\,\,\,\rho^{-1}(\mu a -b) - 1\\
-\epsilon
\end{pmatrix}\\
d_{0}(A) &= \left(\mu a - b -\rho \quad \mu c -d -\zeta\rho \quad \epsilon^T\Omega\right)
\end{align}

For the antisymmetric one, there is an extra $\omega^{-1}$ in $d_2$ to raise the $\epsilon^T$ indices
\begin{align}
d_{-1}(A) &= 
\begin{pmatrix}
-\rho^{-1}(\mu c - d) + \zeta \\
\,\,\,\,\,\rho^{-1}(\mu a -b) - 1\\
-\epsilon
\end{pmatrix}\\
d_{0}(A) &= \left(\mu a - b -\rho \quad \mu c -d -\zeta\rho \quad \omega^{-1}\epsilon^T\Omega\right)
\end{align}

The symmetry properties of $\rho$, \eqref{eq:rho_syms_1}, \eqref{eq:rho_syms_2}, imply there is an isomorphism between these chain complexes and their duals. We can write this isomorphism as a commuting diagram. Denote as $V_i$ the dual vector space $(V^i)^*$. Then for the adjoint chiral algebra, we find the following diagram commutes
\begin{equation}\label{eq:commuting_diagrams}
\begin{tikzcd}
 V^{-2} \arrow[r,"d_{-1}(A)"] \arrow[d,"K_{-2}"] & V^{-1} \arrow[r,"d_{0}(A)"] \arrow[d,"K_{-1}"] & V^{0} \arrow[d,"K_{0}"] \\
V_0 \arrow[r,"d_0^T(-A)"] & V_{-1} \arrow[r,"d_{-1}^T(-A)"] & V_{-2} 
\end{tikzcd}    
\end{equation}
with
\begin{equation}
K_{-2} = \omega\,; \qquad
K_{-1} =\begin{pmatrix}
   0 & -\omega\rho & 0   \\
\omega\rho &   0   & 0   \\  
   0 &   0   & -\Omega
\end{pmatrix};\qquad
K_{0} = \omega 
\end{equation}

This implies
\begin{align}
(V^\bullet,d(A)) \cong (V_\bullet,d^T(-A)) 
\end{align}
So we see it's compatible with an orientifold; It identifies the Chan-Paton bundle at $A$ with the dual Chan-Paton bundle at $-A$!

For the antisymmetric one, we have instead
\begin{equation}
\begin{tikzcd}
 V^{-2} \arrow[r,"d_{-1}(A)"] \arrow[d,"K_{-2}"] & V^{-1} \arrow[r,"d_0(A)"] \arrow[d,"K_{-1}"] & V^{0} \arrow[d,"K_{0}"] \\
V_{0} \arrow[r,"d_0^T(A)"] & V_{-1} \arrow[r,"d_{-1}^T(A)"] & V_{-2} 
\end{tikzcd}    
\end{equation}
with
\begin{equation}
K_{-2} = \1\,; \qquad
K_{-1} =\begin{pmatrix}
   0 & -\rho & 0   \\
\rho &   0   & 0   \\  
   0 &   0   & \Omega
\end{pmatrix};\qquad
K_{0} = \1 
\end{equation}

Therefore
\begin{align}
(V^\bullet,d(A)) \cong (V_\bullet,d^T(A)) 
\end{align}

from which we conclude, the Chan-Paton bundle is identified with its dual, with no orbifolding involved in the identification.

\section{Conclusions}

In \cite{giant_gravitons_kasia}, the authors were able to construct, from correlators of determinant operators in the $U(N)$ chiral algebra, the corresponding dual brane in the bulk. In this work, we improved upon their construction by expressing the data of the brane in terms of a complex of coherent sheaves, a more natural way to define branes in the B-model. 
For correlators  mixing determinant and baryon operators
\begin{align}\label{eq:instanton_deformations_conclusions}
\int [d\psi d\psibar] e^{\psibar Z\psi + \psibar J \epsilon + \bar\epsilon I\psi}
\end{align}  
our approach generalizes naturally, producing a new derived coherent sheaf. We propose that this sheaf encodes the data of a conjectural dual D-brane system, that mixes the determinant D1 brane and the space-filling D5 brane.

We further extended this construction to chiral algebras with $SO$ and $Sp$ gauge groups in symmetric and antisymmetric representations. The resulting brane systems exhibit $\bZ_2$ identifications consistent with conjectures of the holographic dual theories \cite{Jake_Abajian:2024rjq}. Specifically, for chiral algebras in the adjoint representation, the holographic dual is conjectured to be Kodaira-Spencer theory on $SL_2(\mathbb{C})/\mathbb{Z}_2$. Our candidate dual brane naturally incorporates the required orientifold identifications.  

For chiral algebras where $X$ and $Y$ lie in the antisymmetric representation of $Sp(2N)$ or the symmetric representation of $SO(N)$, the dual theory is conjectured to be the Type I topological string on $SL_2(\mathbb{C})$. In this case as well, the Chan-Paton bundle of the candidate dual brane possesses the identifications necessary to match the dual theory. 

\section{Acknowledgments}

I am grateful to Davide Gaiotto for suggesting the problem to me and to Kevin Costello and Jacob Abajian for insightful discussions about the dual theories of these chiral algebras. I also want to thank Kasia Budzik for her invaluable help during the early stages of this work.

This research was supported in part by a grant from the Krembil Foundation. This is work is supported by the NSERC Discovery Grant program and by the Perimeter Institute for Theoretical Physics. Research at Perimeter Institute is supported in part by the Government of Canada through the Department of Innovation, Science and Economic Development Canada and by the Province of Ontario through the Ministry of Colleges and Universities.

% I want to thank Davide Gaiotto for suggesting the main ideas of this work, and Kevin Costello and Jacob Abajian for pointing out details of the corresponding dual theories of these chiral algebras. I also want to thank Kasia Budzik for her invaluable help during the early stages of this work.  This research was supported in part by a grant from the Krembil Foundation. This is work is supported by the NSERC Discovery Grant program and by the Perimeter Institute for Theoretical Physics. Research at Perimeter Institute is supported in part by the Government of Canada through the Department of Innovation, Science and Economic Development Canada and by the Province of Ontario through the Ministry of Colleges and Universities.

\appendix

\section{In detail calculation of \texorpdfstring{$SO(N)$ and $Sp(N)$}{SO(N) and Sp(N)} determinants}\label{app:detailed_SO_Sp_calculation}

In this appendix, we show the in-detail calculation of expectation values of products of determinant operators, using the Hubbard-Stratonovich transformation and the saddle point approximation. We'll do this for both gauge groups $SO(N)$ and $Sp(N)$, and for both $Z$ symmetric and antisymmetric for each gauge group. The OPE's for the theories in question are
\begin{equation}\label{eq:Z_OPE}
    Z(z ; u)^{ab} Z(z' ; u')^{cd} \sim \begin{cases}
    -\dfrac{1}{N}\dfrac{u - u'}{z - z'}\;\dfrac{\delta^{ac}\delta^{bd} \pm \delta^{ad}\delta^{bc}}{2} & \text{for $SO(N)$}\\[9pt]
    \,\,\,\,\dfrac{1}{N}\dfrac{u - u'}{z - z'}\; \dfrac{\Omega^{ac}\Omega^{bd} \pm \Omega^{ad}\Omega^{bc}}{2} & \text{for $Sp(N)$}
    \end{cases}
\end{equation}
where $\pm$ is plus for $Z$ symmetric and minus when antisymmetric. We also chose a different sign in front of the OPE for each gauge group. This decision is merely cosmetic, as it guarantees the expression for saddle point equation we find below are the same for all chiral algebras.

The determinant operators are constructed from fields $\psi^a_i$ with an extra index $i$. As mentioned in Section \ref{subsec:dets_SO_Sp}, we interpret this to mean that the determinant brane is now a stack with, say, $k$ branes and where $i$ is its Chan-Paton index. In very explicit detail, the determinant operators we study in this appendix are of the form
\begin{align}
&\int [d\psi] \,\exp\biggl(\frac{1}{2}\psi_{ai}m^{ij}\delta^{ab}\psi_{bj} + \frac{1}{2}\psi_{ai}Z^{ab}P^{ij}\psi_{bj}\biggr)\,\, \text{ for gauge group $SO$}\\
&\int [d\psi] \,\exp\biggl(\frac{1}{2}\psi_{ai}m^{ij}\Omega^{ab}\psi_{bj} + \frac{1}{2}\psi_{ai}Z^{ab}P^{ij}\psi_{bj}\biggr) \text{ for gauge group $Sp$}
\end{align}

% \begin{align}
% \int [d\psi] \,\exp\biggl(\frac{1}{2}\psi_{ai}m^{ij}\delta^{ab}\psi_{bj} + \frac{1}{2}\psi_{ai}Z^{ab}\delta^{ij}\psi_{bj}\biggr)&\text{ for gauge group } SO \text{ and $Z$ antisymmetric}\\
% \int [d\psi] \,\exp\biggl(\frac{1}{2}\psi_{ai}m^{ij}\Omega^{ab}\psi_{bj} + \frac{1}{2}\psi_{ai}Z^{ab}\Omega^{ij}\psi_{bj}\biggr) &\text{ for gauge group } Sp \text{ and $Z$ symmetric}\\
% \int [d\psi] \,\exp\biggl(\frac{1}{2}\psi_{ai}m^{ij}\delta^{ab}\psi_{bj} + \frac{1}{2}\psi_{ai}Z^{ab}\delta^{ij}\psi_{bj}\biggr)&\text{ for gauge group } SO \text{ and $Z$ symmetric}\\
% \int [d\psi] \,\exp\biggl(\frac{1}{2}\psi_{ai}m^{ij}\Omega^{ab}\psi_{bj} + \frac{1}{2}\psi_{ai}Z^{ab}\Omega^{ij}\psi_{bj}\biggr) &\text{ for gauge group } Sp \text{ and $Z$ antisymmetric}
% \end{align}

where $m^{ij}$ is antisymmetric (respectively, symmetric) for the gauge group $SO$ (respectively, $Sp$). Similarly, for $Z$ symmetric, the bilinear form $P^{ij}$ must be antisymmetric and we set it to the symplectic form $\Omega^{ij}$ without loss of generality. In the case of $Z$ antisymmetric, we instead set $P^{ij} = \delta^{ij}$. 
We will treat all cases in parallel by expressing all of the above as
\begin{align}
\int [d\psi] \,\exp\biggl(\frac{1}{2}\psi_{ai}m^{ij}\tsr{\psi}{^a_j} + \frac{1}{2}\psi_{ai}Z^{ab}\tsr{\psi}{_b^i}\biggr)
\end{align}

However, one must keep careful track of which bilinear form one is using to raise and lower the different indices.

Now we proceed to calculate expectation value of their product:

\begin{align}\label{eq:prods_of_dets_SO_Sp}
\biggl< \prod_\alpha :\int [d\psi_\alpha] \,\exp\biggl(\frac{1}{2}\psi_{\alpha ai}\tsr{m}{_\alpha^{ij}}\tsr{\psi}{_\alpha^a_j} + \frac{1}{2}\psi_{\alpha ai}Z_\alpha^{ab}\tsr{\psi}{_\alpha_b^i}\biggr): \biggr>
\end{align}
% \biggl<\prod_\alpha \mathcal{O}_\alpha\biggr> = \
where  $Z_\alpha\coloneqq Z(z_\alpha,u_\alpha)$.

\subsection{Wick Contracting the \texorpdfstring{$Z$}{Z} fields}

The first step is to perform the wick contractions of the $Z$ fields in Eq \eqref{eq:prods_of_dets_SO_Sp}. To do so, we rewrite the interaction between the $Z$ and $\psi$ fields in Eq. \eqref{eq:prods_of_dets_SO_Sp} as:

\begin{equation}
Z_\alpha^{ab} \tensor{J}{^\alpha_a_b} %  + \frac{\tensor{\psi}{_\alpha_a} \tensor{m}{_\alpha} \tensor{\psi}{_\alpha^a}}{2}
\end{equation}

where
\begin{equation}\label{eq:J}
    J_{\alpha a b} \coloneqq \frac{\tensor{\psi}{_\alpha_a_i} \tensor{\psi}{_\alpha_b^i}}{2}
\end{equation}

Wick contracting the $Z$'s in these products of exponentials we get:

\begin{align}
\biggl<\prod_\alpha:&\exp\left(Z_\alpha^{ab} J^\alpha_{ab} + \frac{\tensor{\psi}{_\alpha_a_i} \tensor{m}{_\alpha^{ij}} \tensor{\psi}{_\alpha^a_j}}{2}\right): \biggr>=\\
&= \exp\left(\sum_{\alpha \neq \beta} \frac{1}{2}J^\alpha_{ab} \left\langle Z_\alpha^{ab}  Z_\beta^{cd} \right\rangle J^\beta_{cd} + \sum_\alpha\frac{\tensor{\psi}{_\alpha_a_i} \tensor{m}{_\alpha^{ij}} \tensor{\psi}{_\alpha^a_j}}{2}\right)\nonumber\\
&\;\;\vdots\nonumber\\
&= \exp\left(\mp\frac{1}{2N} \sum_{\alpha\neq\beta} \frac{z_\alpha -z_\beta}{u_\alpha - u_\beta}J^\alpha_{ab} J^{\beta a b} + \sum_\alpha \frac{\tensor{\psi}{_\alpha_a_i} \tensor{m}{_\alpha^{ij}} \tensor{\psi}{_\alpha^a_j}}{2}\right) 
\end{align}
where the $-$ is for the $SO$ chiral algebras and $+$ for the $Sp$ ones.

Defining $d_{\alpha\beta} \coloneqq \frac{z_\alpha -z_\beta}{u_\alpha - u_\beta}$ and plugging back the expression for $J$ from Eq \eqref{eq:J}, we conclude that the problem of calculating expectation values of determinant operators reduces to the problem of calculating the partition function of a fermionic theory:
% \left\langle\prod_\alpha :\pf( Z_\alpha +m_\alpha):\right\rangle = 
\begin{equation}
\int [\dd\psi\dd\psibar] \,e^{S_F}
\end{equation}

with $S_F$ is defined as

\begin{equation}
    S_F \coloneqq \mp\frac{1}{2N}\sum_{\alpha\neq\beta} d_{\alpha\beta}\frac{\tensor{\psi}{_\alpha_a_i}  \tensor{\psi}{_\alpha_b^i}}{2} \frac{\tensor{\psi}{_\beta^a_j}\tensor{\psi}{_\beta^b^j}}{2} + \sum_\alpha \frac{\tensor{\psi}{_\alpha_a_i} \tensor{m}{_\alpha^{ij}} \tensor{\psi}{_\alpha^a_j}}{2}
\end{equation}

To remove the clutter caused by the myriad of indices of the quartic term, we introduce a couple definitions. First, we leave the contraction of the gauge indices $a/b$ implicit by defining a bilinear product $(\cdot)$
\begin{align}
\mp\frac{1}{2N}\sum_{\alpha\neq\beta} d_{\alpha\beta}\frac{\tensor{\psi}{_\alpha_a_i}  \tensor{\psi}{_\alpha_b^i}}{2} \frac{\tensor{\psi}{_\beta^a_j}\tensor{\psi}{_\beta^b^j}}{2} &= \mp\frac{1}{2N}\sum_{\alpha\neq\beta} d_{\alpha\beta}\left(-\frac{\tensor{\psi}{_\alpha_a_i}\tensor{\psi}{_\beta^a_j}}{2} \frac{\tensor{\psi}{_\alpha_b^i}\tensor{\psi}{_\beta^b^j}}{2} \right)\nonumber\\
&= \pm\frac{1}{2N}\sum_{\alpha\neq\beta} d_{\alpha\beta}\frac{(\tensor{\psi}{_\alpha_i}\cdot\tensor{\psi}{_\beta_j})}{2} \frac{(\tensor{\psi}{_\alpha^i}\cdot\tensor{\psi}{_\beta^j})}{2}\nonumber\\
&= -\frac{1}{2N}\sum_{\alpha\neq\beta} d_{\alpha\beta}\frac{(\tensor{\psi}{_\alpha_i}\cdot\tensor{\psi}{_\beta_j})}{2} \frac{(\tensor{\psi}{_\beta^j}\cdot\tensor{\psi}{_\alpha^i})}{2} \label{eq:manipulating_quartic_fermions}
\end{align}

The definition of $(\cdot)$ naturally depends on the gauge group under consideration:
\begin{align}\label{eq:cdot_def}
    \psi \cdot \psi \coloneqq \psi_a\psi^a = \begin{cases}
        \psi_a \delta^{ab} \psi_b \qquad& \text{for SO}\\
        \psi_a \Omega^{ab} \psi_b \qquad& \text{for Sp}
    \end{cases}
\end{align}

The second notation change will be to define new indices $I, J$ that subsume both the $i,j$ and the $\alpha, \beta$ indices:
\begin{equation}
    I \coloneqq (i,\alpha), \qquad J \coloneqq (j,\beta)
\end{equation}

So that the fermionic action takes the form:
\begin{align}\label{eq:fermion_action}
    S_F &= -\frac{1}{2N}\sum_{\substack{I,J \\ \alpha\neq\beta}} d_{IJ}\frac{\psi_I\cdot\psi_J}{2} \frac{\psi^J\cdot\psi^I}{2} + \sum_\alpha \frac{\tensor{\psi}{_\alpha_a_i} \tensor{m}{_\alpha^{ij}} \tensor{\psi}{_\alpha^a_j}}{2} \\
\end{align}

In doing this change one must keep in mind that, 1) the summation over $I, J$ is s.t. they have $\alpha\neq\beta$, and 2) that $d_{IJ} = d_{\alpha\beta}$. \\

% Before proceeding to perform the Hubbard-Stratonovich transformation, it is worth mentioning that the $(\cdot)$ operation we just defined turns out to be anti-symmetric for SO and symmetric for Sp; this will end up defining the symmetry properties of the indices of the bosonic fields that will be introduced in the following section.

\subsection{Applying the Hubbard-Stratonovich transformation}

By integrating in a $\rho$ field into the action, we may remove the quartic``$\psi$-$\psi$" interaction and re-express it as an interaction mediated through a bosonic $\rho$ field: 
\begin{align}
S_F \rightarrow S_{HS} &= \sum_{\substack{I,J \\ \alpha\neq\beta}} \frac{N}{d_{IJ}} \frac{\rho_{IJ}\rho^{JI}}{2} - \frac{d_{IJ}}{2N}\frac{(\psi_I \cdot \psi_J)}{2} \frac{(\psi^J \cdot \psi^I)}{2}  + \sum_\alpha \frac{\tensor{\psi}{_\alpha_a_i} \tensor{m}{_\alpha^{ij}} \tensor{\psi}{_\alpha^a_j}}{2} 
\end{align}

which after making the shift $\rho_{IJ} \mapsto \rho_{IJ} + \frac{d_{IJ}}{N}\frac{(\psi_{I} \cdot \psi_J)}{2}$, becomes: 
\begin{align}
 S_{HS} &\rightarrow  \sum_{\substack{I\,J\\\alpha\neq\beta}}  \frac{N}{d_{IJ}} \dfrac{(\rho_{IJ} + d_{IJ}\frac{(\psi_{I} \cdot \psi_J)}{2N})(\rho^{JI} + d_{JI}\frac{(\psi^{J} \cdot \psi^I)}{2N})}{2} - \frac{d_{IJ}}{2N}\frac{(\psi_I \cdot \psi_J)}{2} \frac{(\psi^J \cdot \psi^I)}{2}  +\nonumber\\
&\qquad\qquad\qquad\qquad\qquad\qquad\qquad\qquad\qquad\qquad   + \sum_\alpha \frac{\tensor{\psi}{_\alpha_a_i} \tensor{m}{_\alpha^{ij}} \tensor{\psi}{_\alpha^a_j}}{2} \nonumber\\
 &= \sum_{\substack{I\,J\\\alpha\neq\beta}} \frac{N}{d_{IJ}} \frac{\rho_{IJ}\rho^{JI}}{2} + \sum_{\substack{I\,J\\\alpha\neq\beta}} \rho^{IJ}\;\frac{\psi_I \cdot \psi_J}{2} + \sum_\alpha \frac{\tensor{\psi}{_\alpha_a_i} \tensor{m}{_\alpha^{ij}} \tensor{\psi}{_\alpha^a_j}}{2}\nonumber\\
 &= \sum_{\substack{I\,J\\\alpha\neq\beta}} \frac{N}{d_{IJ}} \frac{\rho_{IJ}\rho^{JI}}{2} + \sum_{I,J} \rho^{IJ}\;\frac{\psi_I \cdot \psi_J}{2}
\end{align}

Where in the last line we absorbed the mass term into the definition of $\rho$ by setting:
\begin{equation}
    \tensor{\rho}{^\alpha^i^\alpha^j} \coloneqq \tensor{m}{_\alpha^{ij}}
\end{equation}

for those $I, J$ that have $\alpha=\beta$.\\

Observe that the matrix $\rho_{IJ}$ has different symmetry properties for each gauge group. These are inherited from the symmetry properties of the $(\cdot)$ product defined in Eq. \eqref{eq:cdot_def} and from the fact that the $\psi$'s anti-commute:
\begin{align}\label{eq:rho_symmetries}
    \rho_{IJ} &= - \rho_{JI} \quad\text{ for } SO\\
    \rho_{IJ} &= + \rho_{JI} \quad\text{ for } Sp
\end{align}

\subsection{Integrating out the fermions}

\subsubsection{integrating out the fermions for SO(N)}
To integrate out the fermions, we rewrite the interaction terms of the new action:
\begin{align}
(S_{HS})_{\text{int}} &= \rho^{IJ} \frac{\psi_I \cdot \psi_J}{2} \nonumber\\
&= \frac{1}{2}\;\psi_{Ia}\rho^{IJ}\delta^{ab} \psi_{Jb}\nonumber\\
&= \frac{1}{2}\;\psi^T (\rho\otimes \1)\psi
\end{align}

So that, integrating out the fermions we get:
\begin{align}
\int [\dd \psi\dd\psibar] e^{(S_{HS})_{\text{int}}} &= \pf\left(\rho\otimes\mathds{1}\right) \nonumber\\
&= \det\left(\rho\otimes\mathds{1}\right)^{\frac{1}{2}}\nonumber\\
&= \det(\rho)^{\frac{N}{2}}\nonumber\\
&= e^{\frac{N}{2}\Tr\ln(\rho)}
\end{align}

\subsubsection{Integrating out the fermions for Sp(N)}

To integrate out the fermions, we rewrite the interaction terms of the new action as we did for SO(N):

\begin{align}
(S_{HS})_{\text{int}} &= -\rho^{IJ} \frac{\psi_I \cdot \psi_J}{2} \nonumber\\
&= -\frac{1}{2}\;\psi_{Ia}\rho^{IJ}\Omega^{ab} \psi_{Jb}\nonumber\\
&= -\frac{1}{2}\;\psi^T (\rho\otimes \Omega)\psi
\end{align}

So that, integrating out the fermions we get:

\begin{align}
\int [\dd \psi\dd\psibar] e^{(S_{HS})_{\text{int}}} &= \pf\left(\rho\otimes\Omega\right) \nonumber\\
&= \det\left(\rho^2\otimes\cancelto{-\1}{\Omega^2}\right)^{\frac{1}{4}}\nonumber\\
&= (\cancel{(-1)^{k N}})^{\frac{1}{4}}\det(\rho)^{\frac{N}{2}}\nonumber\\
&= e^{\frac{N}{2}\Tr\ln(\rho)}
\end{align}

Where in going from the first to the second line we raised the Pfaffian to the fourth power and took a fourth root, and in the third line we canceled the ${(-1)^{k N}}$ factor because $N$ is even for $Sp(N)$.

\subsection{The Final \texorpdfstring{$\rho$}{\textrho} Action}

% The final action for the $\rho$ field becomes
% \begin{align}
%     S_\rho &= N \sum_{\substack{I, J\\\alpha\neq\beta}} -\frac{\rho_{IJ}\rho^{IJ}}{2d_{IJ}} + \frac{N}{2}\Tr\ln(\rho)
% \end{align}

We conclude that the problem of calculating the expectation value of products of determinant operators reduces to the problem of calculating the integral:

\begin{align}
\biggl< \prod_\alpha :\int [d\psi_\alpha] \,\exp\biggl(\frac{1}{2}\psi_{\alpha ai}\tsr{m}{_\alpha^{ij}}\tsr{\psi}{_\alpha^a_j} + \frac{1}{2}\psi_{\alpha ai}Z_\alpha^{ab}\tsr{\psi}{_\alpha_b^i}\biggr): \biggr> = \frac{1}{\mathcal{Z}_\rho}\int[\dd\rho] e^{S_\rho}
\end{align}

where
\begin{align}
\mathcal{Z}_\rho &\coloneqq \int[\dd\rho] \exp\left(N\sum_{\substack{I, J\\\alpha\neq\beta}}\frac{\rho_{IJ}\rho^{JI}}{2d_{IJ}}\right)\\
S_\rho &\coloneqq N \sum_{\substack{I, J\\\alpha\neq\beta}} \frac{\rho_{IJ}\rho^{JI}}{2d_{IJ}} + \frac{N}{2}\Tr\ln(\rho)\label{eq:final_action_without_indices}
\end{align}

We rescale $\rho \mapsto \frac{1}{\sqrt{2}} \rho$ to make the final saddle equation look the same as the on for $U(N)$ chiral algebras. After this cosmetic change, the action for $\rho$ (modulo an unimportant additive constant coming from the log) becomes
\begin{align}
S_\rho &\coloneqq N \sum_{\substack{I, J\\\alpha\neq\beta}}\frac{\rho_{IJ}\rho^{JI}}{4d_{IJ}} + \frac{N}{2}\Tr\ln(\rho)\label{eq:final_action_without_indices_0}
\end{align}

and in the large $N$ limit we arrive at the saddle equation
\begin{align}
(z_\alpha - z_\beta)\rho_{\alpha i \beta j} + (u_\alpha - u_\beta)(\rho^{-1})_{\alpha i \beta j} = 0
\end{align}

Despite satisfying the same saddle point equation, what distinguishes the gauge group and the representation for $Z$ we started from is
\begin{enumerate}
\item The symmetry properties of $\rho$:
\begin{align}\label{eq:rho_symmetries_again}
    \rho_{\alpha i \beta j} &= - \rho_{\beta j\alpha i} \quad\text{ for } SO\\
    \rho_{\alpha i \beta j} &= + \rho_{\beta j\alpha i} \quad\text{ for } Sp
\end{align}

\item What bilinear form we use to raise the $i,j$ indices:
\begin{align}
\Omega^{ij} &\text{ for $Z$ in the symmetric representation}\\
\delta^{ij} &\text{ for $Z$ in the antisymmetric representation}
\end{align}

\end{enumerate}

\subsection{Including Flavors}

Adding fermionic flavors to our chiral algebra with OPE
\begin{align}
I^a_A(z) I^b_B(z') \sim 
\begin{cases}
\frac{1}{z-z'} \delta^{ab}\,\delta_{AB} \text{ for gauge group $SO$}\\
\frac{1}{z-z'} \Omega^{ab}\,\Omega_{AB} \text{ for gauge group $Sp$}
\end{cases}
\end{align}

The determinant/baryon correlators have the form
\begin{align}\label{eq:prods_of_det_bar_SO_Sp}
\biggl< \prod_\alpha :\int [d\psi_\alpha] \,\exp\biggl(\frac{1}{2}\psi_{\alpha ai}\tsr{m}{_\alpha^{ij}}\tsr{\psi}{_\alpha^a_j} + \frac{1}{2}\psi_{\alpha ai}Z_\alpha^{ab}\tsr{\psi}{_\alpha_b^i} + \tsr{\psi}{_\alpha^i_a} I^a_A(z_\alpha) \tsr{\epsilon}{_\alpha^A_i}\biggr): \biggr>
\end{align}

where the $a$ indices in $\psi_a I^a$ are contracted with $\delta^{ab}$ or $\Omega^{ab}$ depending on the gauge group.

Performing the wick contractions of the $I$ fields adds to Eq \eqref{eq:fermion_action} the term
\begin{align}
\sum_{\alpha\neq\beta}\frac{1}{2} \tsr{\psi}{_\alpha_a^i}\tsr{\psi}{_\beta_b^j}\tsr{\epsilon}{_\alpha^A_i}\tsr{\epsilon}{_\beta^B_j}\bigl<I^a_A I^b_B \bigr>
\end{align}
which for both $SO(N)$ and $Sp(N)$ gauge group equals
\begin{align}
&\sum_{\alpha\neq\beta}\frac{1}{2} \frac{1}{z_\alpha - z_\beta}\tsr{\psi}{_\alpha_a^i}\tsr{\psi}{_\beta^a^j}\tsr{\epsilon}{_\alpha^A_i}\tsr{\epsilon}{_\beta_A_j}
\end{align}

Once we integrate out the fermions we arrive at the new action for $\rho$
\begin{align}
S_\rho = N \sum_{\substack{I, J\\\alpha\neq\beta}}\frac{\rho_{IJ}\rho^{JI}}{4d_{IJ}} + \frac{N}{2}\Tr\ln\left(\rho_{IJ} + \frac{1}{z_I-z_J}\epsilon_{IA}\tsr{\epsilon}{_J^{A}}\right)
\end{align}

which after shifting $\rho_{IJ} \mapsto \rho_{IJ} -\frac{1}{z_I-z_J}\epsilon_{IA}\tsr{\epsilon}{_J^{A}}$, leads to the saddle equation
\begin{align}
(z_\alpha - z_\beta)\rho^{\alpha i \beta j} + (u_\alpha - u_\beta)(\rho^{-1})_{\alpha i \beta j} = \epsilon_{\alpha iA}\tsr{\epsilon}{_\beta _j^{A}}
\end{align}

\bibliographystyle{JHEP}
\bibliography{mono}

\end{document}